\documentclass[final,3p,times]{elsarticle}

\usepackage{lineno,hyperref}
\usepackage{amssymb}
\usepackage{amsmath}
\usepackage{amsfonts}
\usepackage{graphicx}
\usepackage{array}
\usepackage{booktabs}
\usepackage{float}
\usepackage{multirow}
\usepackage{color}
\usepackage[ruled]{algorithm2e}
\usepackage{subfigure}
\usepackage{bbding}
\usepackage{hyperref}
\usepackage{diagbox}
\usepackage{hyperref}
\usepackage{amsfonts,amssymb} 
\usepackage{soul, color, xcolor}
\usepackage{caption} 

\hypersetup{hidelinks,
	colorlinks=true,
	allcolors=black,
	pdfstartview=Fit,
	breaklinks=true}
\modulolinenumbers[5]
\biboptions{sort&compress}
\journal{Journal of \LaTeX\ Templates}

\begin{document}  

\begin{frontmatter}

\title{A Comprehensive Survey on Magnetic Resonance Image Reconstruction
}

\author[mymainaddress]{Xiaoyan Kui}
\author[mymainaddress]{Zijie Fan}
\author[mymainaddress]{Zexin Ji\corref{mycorrespondingauthor}}
\cortext[mycorrespondingauthor]{Co-Corresponding Authors}
\ead{zexin.ji@csu.edu.cn}
\author[mysecondaddress]{Qinsong Li\corref{mycorrespondingauthor}}\ead{qinsli.cg@csu.edu.cn}
\author[mymainaddress]{Chengtao Liu}
\author[mythirdaddress]{Weixin Si}
\author[mymainaddress]{Beiji Zou}

\address[mymainaddress]{School of Computer Science and Engineering, Central South University, Changsha, 410083, China}
\address[mysecondaddress]{Big Data Institute, Central South University, Changsha, 410083, Hunan, China}
\address[mythirdaddress]{Shenzhen Institute of Advanced Technology, Chinese Academy of Science, Shenzheng, 518055, Guangdong, China}

\begin{abstract}

Magnetic resonance imaging (MRI) reconstruction is a fundamental task aimed at recovering high-quality images from undersampled or low-quality MRI data. This process enhances diagnostic accuracy and optimizes clinical applications. In recent years, deep learning-based MRI reconstruction has made significant progress. Advancements include single-modality feature extraction using different network architectures, the integration of multimodal information, and the adoption of unsupervised or semi-supervised learning strategies. However, despite extensive research, MRI reconstruction remains a challenging problem that has yet to be fully resolved. This survey provides a systematic review of MRI reconstruction methods, covering key aspects such as data acquisition and preprocessing, publicly available datasets, single and multi-modal reconstruction models, training strategies, and evaluation metrics based on image reconstruction and downstream tasks. Additionally, we analyze the major challenges in this field and explore potential future directions.

\end{abstract}


\end{frontmatter}

\section{Introduction}\label{1.Introduction}

Magnetic Resonance Imaging (MRI) is a widely used non-invasive imaging technique in the medical field. Compared with X-ray or Computed Tomography (CT), MRI excels in soft tissue imaging, providing richer pathological information for clinical diagnosis, especially for the detection of brain tissue, joints, and internal organs. However, the long imaging time required for MRI can lead to patient discomfort and significantly increase the probability of motion artifacts. These motion artifacts degrade the quality of MRI images and affect the accuracy of diagnoses made by clinicians. Therefore, accelerating MRI acquisition is crucial for improving clinical practice. To accelerate MRI acquisition, undersampling is commonly employed, which involves collecting only a portion of the k-space raw data. Although the undersampling technique can effectively shorten the scan duration, it significantly degrades the image quality. 

In recent decades, various MR image reconstruction techniques have been developed to reconstruct high-quality MRI images from undersampled data and enhance the speed of MRI scans. Early progress was made through the use of zero-filling and interpolation techniques, which fill in missing data with zeros or estimate values based on neighboring points. While these methods can help address undersampling, they may not fully capture the underlying details and structures of MR images. To improve upon this, parallel imaging methods make use of multiple receiver coils, which accelerate image acquisition and reduce artifacts from undersampling. Compressed sensing, another powerful technique, leverages data sparsity to reconstruct high-quality images from fewer samples. These traditional MRI reconstruction methods often fail to achieve the desired reconstruction quality in highly undersampled spaces. 

Deep learning-based reconstruction methods with their powerful feature extraction and learning capabilities have been able to effectively learn the mapping from undersampled spaces to fully sampled spaces. The design of different network architectures, such as Transformer, diffusion models, deep unfolding networks, and Mamba, has further enhanced performance by optimizing the learning process, reducing artifacts, and enabling more accurate and efficient image reconstruction. This has significantly improved reconstruction performance. 
Enhanced reconstruction quality in MRI plays a vital role in improving clinical tasks. Higher-quality images allow clinicians to detect abnormalities more reliably, identify early-stage diseases, and differentiate between benign and malignant growths. Detailed scans also aid in precise treatment planning, such as in radiation therapy or surgery, where understanding the exact location of tumors is critical. Additionally, high-quality MRI scans improve the monitoring of chronic conditions, enabling more sensitive tracking of changes over time and allowing for timely adjustments in treatment plans. This all contributes to better patient care and outcomes.

\begin{figure}[t]  
    \centering
    \includegraphics[width=1\textwidth]{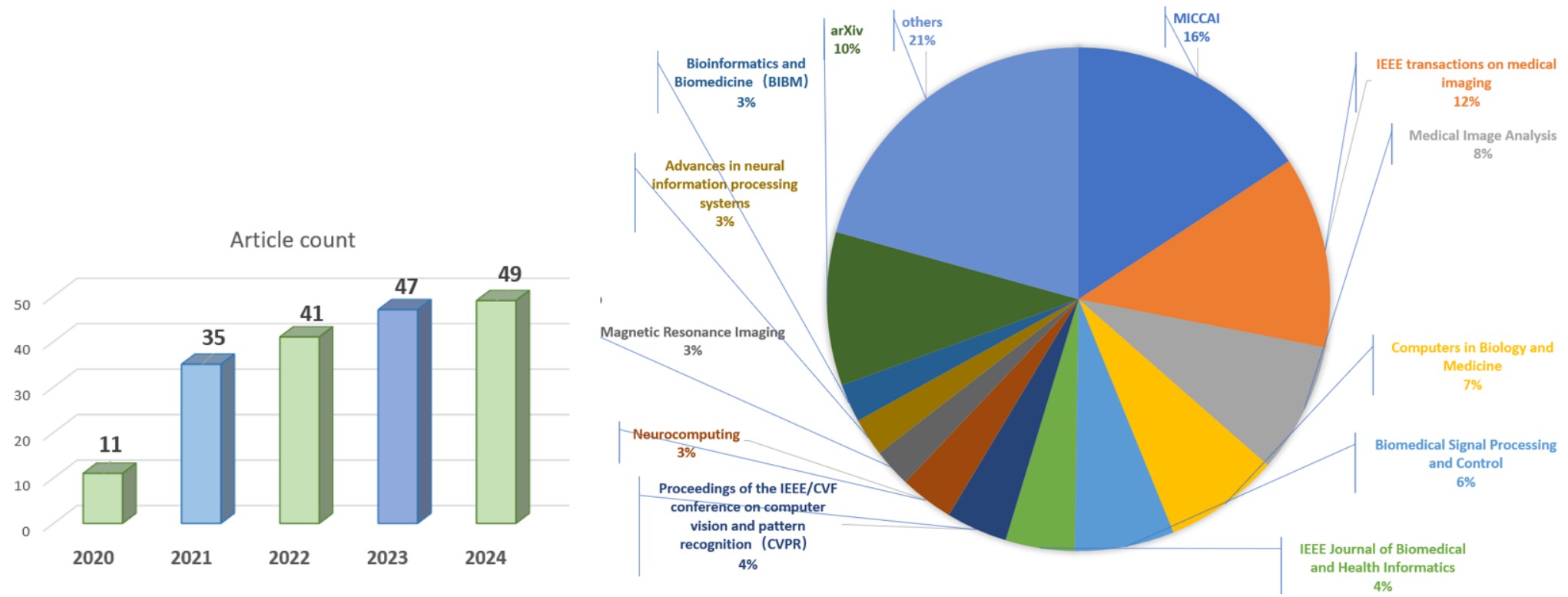}  
    \caption{This survey paper presents an analysis of the articles under investigation. The left section features a bar chart depicting the distribution of paper counts across different years; it is evident that the majority of the reviewed papers were introduced over the past five years. The right section provides a visual representation of the origins of the examined articles, highlighting that our survey encompasses a range of sources pertinent to the domain of MRI reconstruction.}
    \label{fig:paper-classfiy}
    \end{figure}

As shown in Figure~\ref{fig:paper-classfiy}, the number of published articles on deep learning-based MR image reconstruction has shown a significant upward trend over the years. Given the rapid development of deep learning techniques and their application to MRI reconstruction, a thorough review is necessary to summarize the current state of research. Therefore, we provide a brief description of traditional reconstruction methods and conduct a comprehensive review of recent research on deep learning-based MRI reconstruction methods, hoping to provide a reference for MRI imaging research.

The main contributions of this paper are as follows:

\begin{itemize}
    \item Compared with previous review papers on MRI reconstruction~\cite{zeng2021review}, this survey incorporates the latest literature, provides a systematic review of MRI reconstruction methods, analyzes datasets, network architectures, training strategies, loss functions, and evaluation methods, and emphasizes the application value and significance of MRI reconstruction in clinical practice.
    
    \item  This survey further discusses the current challenges and limitations in the field of MRI reconstruction and provides an outlook on its future development directions.
\end{itemize}

\textbf{Paper Collection:} We primarily collected papers from Google Scholar and DBLP. By searching for relevant literature in top journals and conferences(e.g., Medical Image Analysis, IEEE Transactions on Medical Imaging, MICCAI, BIBM, etc.) using the keywords "MRI reconstruction", "MR image reconstruction", and "Magnetic Resonance Image Reconstruction". Further, we used keywords such as "multi-contrast" to find reconstruction methods for multi-modal MRI, and keywords like "self-supervised" and "semi-supervised" to locate reconstruction methods employing different training strategies. By examining the references of these papers, we were able to obtain a more comprehensive overview of the related work. We have categorized the collected literature by year and source, as shown in Figure~\ref{fig:paper-classfiy}.

\begin{figure}[t]  
    \centering
    \includegraphics[width=0.5\textwidth]{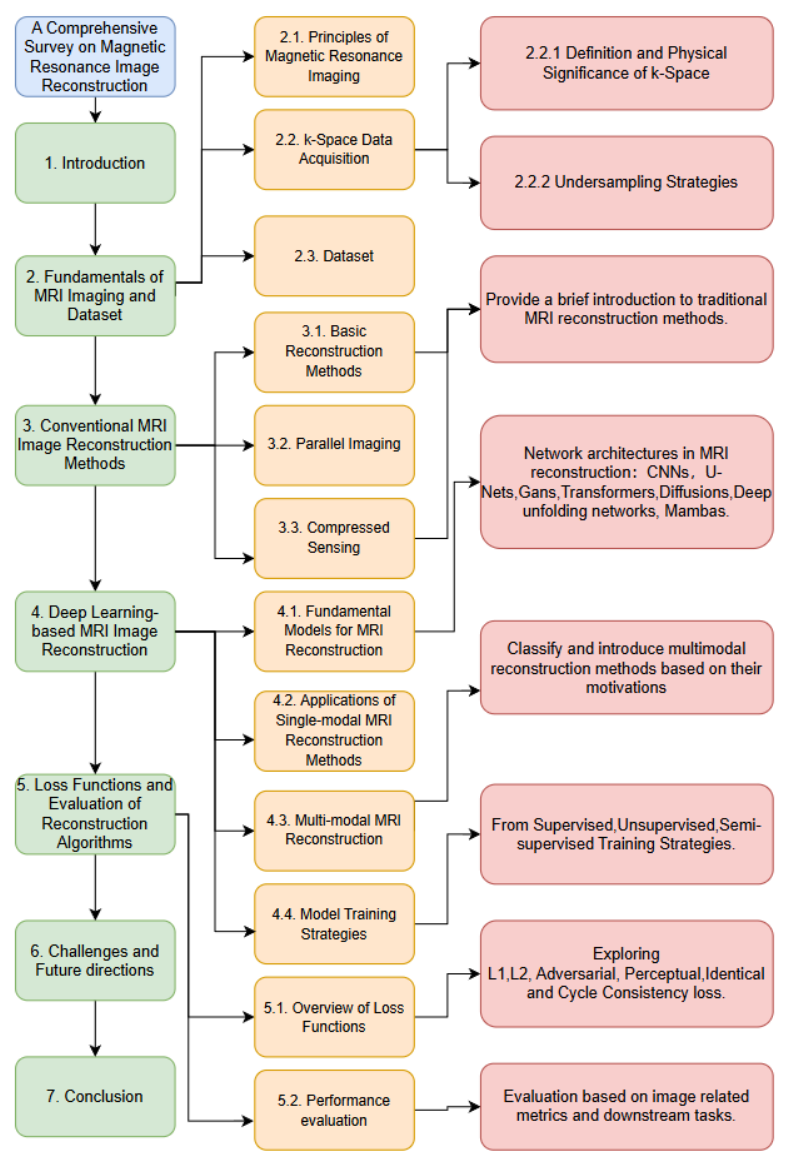}  
    \caption{An overview of the paper organization.}
    \label{fig:overview}
    \end{figure}

\textbf{Paper Organization:} The structure of this paper is shown in Figure~\ref{fig:overview}. Section~\ref{2} provides an overview of the basic principles of MRI imaging (such as undersampling strategies and datasets). Section~\ref{3} introduces traditional MRI reconstruction methods. Section~\ref{4} analyzes deep learning network architectures and their applications in MRI reconstruction. Section~\ref{5} discusses loss functions and evaluation methods for reconstruction quality. Section~\ref{6} addresses the challenges of the collected literature and potential future development directions. Section~\ref{7} concludes this survey.

\section{Fundamentals of MRI Imaging and Dataset}\label{2}
\subsection{Principles of Magnetic Resonance Imaging}~\label{2.1}

Magnetic Resonance Imaging is a non-invasive medical imaging technique. Its fundamental principle relies on the spin behavior of hydrogen protons in an external magnetic field and their response to radiofrequency (RF) pulses. The MRI imaging process consists of several key steps. First, under the influence of a strong magnetic field, the spin orientations of hydrogen protons in the body align, forming a longitudinal magnetization vector. Then, an RF pulse excites the hydrogen protons, causing them to absorb energy and enter a high-energy state, simultaneously generating transverse magnetization. Once the RF pulse is turned off, the protons gradually return to equilibrium, releasing electromagnetic signals. These signals are received by detection coils and stored in k-space, and ultimately, the data from k-space is reconstructed into MRI images using the Inverse Fourier Transform.

\subsection{k-Space Data Acquisition}
\subsubsection{Definition and Physical Significance of k-Space}

In MRI imaging, k-space is used to store the acquired MRI signals in the frequency domain (Fourier domain). Essentially, k-space represents the raw data matrix collected during MRI scanning, which is ultimately transformed into an MRI image via the Inverse Fourier Transform (IFT). From a physical perspective, different regions of k-space encode distinct image information. The central region predominantly stores low-frequency components, which determine the overall image contrast and signal intensity, playing a crucial role in distinguishing tissue gray levels. In contrast, the peripheral regions primarily contain high-frequency components, which represent fine details and edge structures, directly influencing image sharpness and spatial resolution. Therefore, optimizing the distribution of sampled k-space points is essential for achieving high-quality MRI imaging.

To facilitate efficient data acquisition, MRI utilizes gradient fields to regulate the filling of k-space and employs various sampling strategies, such as Cartesian sampling, Radial sampling, and Spiral sampling. However, fully sampling k-space requires prolonged scan times, posing significant limitations in clinical applications, particularly in dynamic imaging and high-resolution scans. To address these challenges, undersampling techniques that leverage the sparsity of k-space have emerged as a pivotal research direction in MRI image reconstruction.

\subsubsection{Undersampling Strategies}

\begin{figure}[t]  
    \centering
    \includegraphics[width=0.6\textwidth]{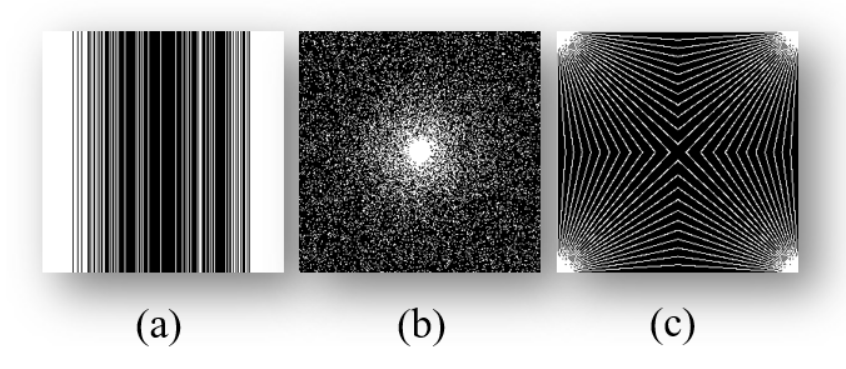}  
    \caption{Some examples of undersampling masks. (a) Cartesian. (b) Gaussian. (c) Radial.}
    \label{fig:masks}
    \end{figure}

In the field of MRI reconstruction, researchers typically utilize specific undersampling masks to fully sampled images to simulate undersampled acquisitions, thus generating the input-label pairs required for training and evaluation.
The design of undersampling masks plays a critical role in determining model training effectiveness and reconstruction quality. 
Based on the distribution of sampled k-space points, undersampling masks can be broadly classified into \textbf{structured sampling} and \textbf{random sampling}. Structured sampling provides greater controllability in traditional Fourier-based reconstruction, whereas random sampling is better suited for deep learning methods. Figure~\ref{fig:masks} displays several undersampling masks.

\textbf{(1) Structured Sampling}  

Structured sampling follows predefined sampling rules, allowing precise control over sampling density. It is commonly used in traditional MRI reconstruction methods based on the Fourier Transform (FT).  

\begin{itemize}  
    \item \textbf{Uniform Sampling:} Sampling points are evenly spaced in k-space, ensuring overall data uniformity. However, periodic data loss due to undersampling may lead to aliasing artifacts.  

    \item \textbf{Equidistant Sampling:} Sampling points are distributed at fixed intervals along different trajectories. This method is commonly used in Radial sampling and Spiral sampling.  

    \item \textbf{Cartesian Sampling:} Sampling points are arranged in a regular grid pattern in k-space. 
\end{itemize}  

\textbf{(2) Random Sampling}  

Random sampling distributes sampled points in k-space in a non-uniform manner, effectively reducing periodic artifacts. This makes it particularly suitable for deep learning-based reconstruction methods.  

\begin{itemize}  
    \item \textbf{Fully Random Sampling:} Sampling points are randomly selected across the entire k-space without a fixed pattern. This approach provides uniform overall coverage but may lead to localized information loss, affecting reconstruction quality.  

    \item \textbf{Gaussian Random Sampling:} Sampling density follows a Gaussian distribution, with denser sampling in the center of k-space and sparser sampling in the periphery. 

    \item \textbf{Poisson Disk Sampling:} By introducing a minimum distance constraint, this method ensures a more uniform distribution of sampling points, preventing excessive clustering.

\end{itemize}

\textbf{ Acceleration factor (AF)} is a key parameter in MRI undersampling, measuring the proportion of k-space data retained compared to fully sampled MRI. For example, AF = 2 indicates that half of the k-space data is discarded, retaining only 50\% of the sampling points, while AF = 4 means that only 25\% of the data is preserved. As the acceleration factor increases, MRI acquisition time is significantly reduced, but this 
also results in the loss of more information.

\subsection{Dataset}

\begin{figure}[t]  
    \centering
    \includegraphics[width=0.45\textwidth]{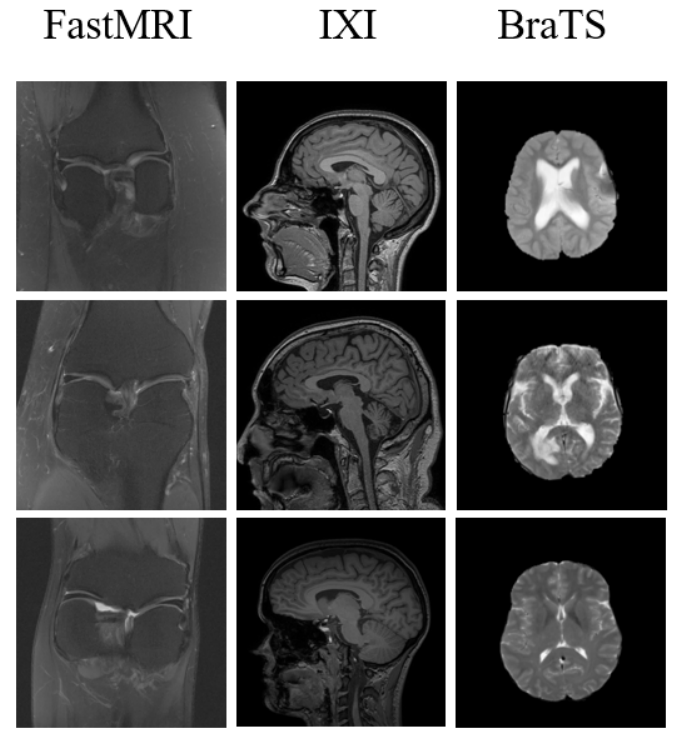}  
    \caption{Some examples of MRI images.}
    \label{fig:data}
    \end{figure}

The following section introduces several commonly used public MRI reconstruction datasets. Figure~\ref{fig:data} shows some examples of MRI images.A summary of publicly available datasets utilized in MRI reconstruction is presented in Table~\ref{tab:dataset}. Here, "Highlights" refers to the main features of each dataset, and "Reference" indicates the literature that utilizes the dataset.

\begin{table}[t]
\centering
\caption{A summary of publicly available datasets utilized in MRI reconstruction.}
    \label{tab:dataset}
\begin{tabular}{@{}cllr@{}}
\toprule
Dataset & Region     & Highlights                                                                                                                                                                                                                                    & Reference \\ \midrule
FastMRI & Brain/Knee & \begin{tabular}[c]{@{}l@{}}Proton Density (PD), Proton Density\\  with Fat Suppression (PD-FS), \\ T1-weighted, T2-weighted, and FLAIR sequences.\\  In addition, it contains both raw\\  MR measurements and clinical MR images\end{tabular} &  \cite{guo2023reconformer,li2024progressive,yi2023frequency,elmas2022federated,sun2025fourier,xuan2022multimodal,arvinte2021deep,guo2021over,hu2021learning,liu2021universal,murugesan2021deep,narnhofer2021bayesian,jun2021joint,li2021multimodal,nitski2020cdf,chen2024fefa,wang2022b,chen2022pyramid,korkmaz2022unsupervised,wang2023dsmenet,ramzi2022nc,rasti2023plug,feng2021deep,aghabiglou2022deep,dong2022invertible,liu2022undersampled,zhou2023rnlfnet,gungor2023adaptive,zheng2023fast,sun2023joint,ekanayake2024mcstra,dar2023parallel,korkmaz2023self,ozturkler2023smrd,zach2023stable,chen2024accelerated,wu2024adaptive,zhao2024center,huang2024noise,wang2024spatial,yan2024cross,chen2024multi,ekanayake2025cl,sriram2020grappanet,jeong2024most}          \\ \cline{3-3}
IXI     & Brain      & \begin{tabular}[c]{@{}l@{}}T1, T2 and PD-weighted, \\ MRA images and Diffusion-weighted image\end{tabular}                                                                                                                                    &  ~\cite{swingan,li2024progressive,elmas2022federated,noor2024dlgan,sheng2024cascade,feng2021task,hu2021self,zhou2023dsformer,kang20243d,chen2022accelerating,chen2024fefa,wei2022undersampled,korkmaz2022unsupervised,feng2021deep,liu2022undersampled,yan2023dc,gungor2023adaptive,lei2023decomposition,yang2023mgdun}         \\  \cline{3-3}
BraTS   & Brain      & \begin{tabular}[c]{@{}l@{}}T1, T1-weighted (T1Gd), \\ T2-weighted, and T2-FLAIR\end{tabular}                                                                                                                                                  &     ~\cite{elmas2022federated,casamitjana2022robust,bian2022learnable,ding2022mri,mittal20243d,lei2023decomposition,yang2023mgdun,chen2024accelerated,sangeetha2024c2,chen2024multi,wei2024misalignment,jeong2024most}      \\ \bottomrule
\end{tabular}
\end{table}

\begin{itemize}
    \item \textbf{FastMRI\footnote{https://fastmri.med.nyu.edu/}}: Released by Facebook AI Research (now Meta AI) in collaboration with NYU Langone Health, fastMRI is one of the most widely used large-scale public datasets for deep learning-based accelerated MRI reconstruction. This dataset includes MRI scans of the knee and brain, providing raw k-space data, multi-coil signals, and magnitude images. It encompasses common MRI contrast types, including Proton Density (PD), Proton Density with Fat Suppression (PD-FS), T1-weighted, T2-weighted, and FLAIR sequences. The knee MRI dataset contains 697 PD-weighted and 701 PD-FS-weighted scans, while the brain MRI dataset includes 382 1.5T T1-weighted, 409 3T T1-weighted, 1655 1.5T T2-weighted, 2524 3T T2-weighted, 126 1.5T FLAIR, and 411 3T FLAIR scans. Notably, each MRI scan in fastMRI consists of multiple 2D slices, meaning that each case comprises a series of sequential 2D MRI images rather than a single 3D volumetric dataset. 

    \item \textbf{IXI\footnote{https://brain-development.org/ixi-dataset/}}: The IXI dataset was collected from three hospitals in London (Hammersmith Hospital, Guy’s Hospital, and the Institute of Psychiatry) and includes nearly 600 fully sampled MRI scans from healthy subjects. This dataset covers multiple MRI contrast types, including T1-weighted, T2-weighted, Proton Density (PD)-weighted, Magnetic Resonance Angiography (MRA), and Diffusion Tensor Imaging (DTI), with DTI data containing diffusion information in 15 directions. Unlike fastMRI, which primarily consists of clinical MRI scans that may include both healthy individuals and patients with pathology, IXI exclusively comprises data from healthy subjects.

    \item \textbf{BraTS\footnote{http://braintumorsegmentation.org/}}: The BraTS (Brain Tumor Segmentation Challenge) dataset is primarily designed for brain tumor segmentation and analysis and is also widely used in MRI image reconstruction research. This dataset comprises MRI scans from patients with various types of brain tumors and includes T1-weighted (T1), contrast-enhanced T1-weighted (T1c), T2-weighted (T2), and Fluid-Attenuated Inversion Recovery (FLAIR) modalities. It provides fully 3D volumetric data, typically stored in NIfTI (.nii.gz) format. The dataset is distinguished by its high resolution, multi-modal information, and comprehensive tumor region annotations. 
    
\end{itemize}

\section{Conventional MRI Image Reconstruction Methods}\label{3}
\subsection{Basic Reconstruction Methods}
To mitigate the loss of k-space information due to undersampling, zero-filling and interpolation techniques are widely employed. Zero-filling is the most straightforward reconstruction strategy, which involves filling unacquired k-space data points with zero values and then applying the inverse fast Fourier transform (IFFT) for image reconstruction. This method is computationally simple and can quickly generate MRI images. However, zero-filling does not introduce any new information essentially; it merely completes the data based on the existing samples. Therefore, under high acceleration ratios with highly undersampled conditions, its ability to recover the image is poor and cannot meet clinical requirements. In contrast, interpolation methods estimate missing k-space data points, thus enhancing the reconstruction of fine image details. Common interpolation techniques include bilinear interpolation, cubic spline interpolation, and Lanczos interpolation. However, when the missing k-space data is extensive, interpolation methods struggle to accurately recover high-frequency components. 

Despite their ability to partially improve undersampled MRI reconstruction, zero-filling and interpolation techniques exhibit inherent limitations, particularly at high acceleration rates, where they fail to effectively restore the missing information. This limitation has driven the development of more advanced reconstruction strategies.

\subsection{Parallel Imaging}
Parallel Imaging (PI)~\cite{griswold2002generalized} is a crucial acceleration technique in the field of MRI. Its fundamental principle is to leverage the spatial sensitivity of multi-channel radiofrequency (RF) coils to reduce the amount of k-space data required for image reconstruction. Traditional MRI acquisition typically relies on a single receiving coil to collect signals, whereas parallel imaging techniques utilize multiple coils simultaneously, each with a distinct spatial sensitivity profile. This allows for undersampling in k-space without introducing severe aliasing artifacts. Consequently, parallel imaging significantly reduces MRI scan time, making it particularly valuable for clinical applications requiring rapid imaging, such as cardiac MRI.

In parallel imaging, two major reconstruction methods are commonly used: SENSE (SENSitivity Encoding)~\cite{ullah2018qr}, which operates in the image domain, and GRAPPA~\cite{inam2022gpu} (GeneRalized Autocalibrating Partially Parallel Acquisitions), which reconstructs images in k-space.
The primary advantage of parallel imaging lies in its ability to increase imaging speed without compromising spatial resolution. However, its main limitation is that the acceleration factor is limited by the number of coils since the acceleration multiple cannot exceed the number of receiving coils.

\subsection{ Compressed Sensing}
With increasing demand for accelerated MRI acquisition, Compressed Sensing (CS) has emerged as a powerful reconstruction framework. In MRI applications, CS assumes that an MRI image is sparse in a certain transform domain, such as wavelet or total variation domains. This allows k-space data to be randomly undersampled, while nonlinear optimization techniques reconstruct the missing information. It is worth noting that the manner in which sparsity is achieved significantly impacts the quality of image reconstruction~\cite{lang2023undersampled}.

CS-based MRI~\cite{lustig2008compressed}reconstruction commonly relies on optimization algorithms such as the Alternating Direction Method of Multipliers (ADMM) and Iterative Shrinkage-Thresholding Algorithm (ISTA)~\cite{zhang2018ista}. ISTA iteratively refines the reconstructed image by applying a sparsity constraint at each iteration, and ISTA is a very common algorithm. There are also some reconstruction methods based on this algorithm that have been proposed~\cite{huang2022ista}.  ADMM enhances computational efficiency by decomposing the optimization problem into subproblems and solving them iteratively.
However, CS-based approaches rely on sparsity assumptions, making them computationally expensive and less effective at very high acceleration rates.

\section{Deep Learning-based MRI Image Reconstruction}\label{4}

Deep learning-based methods leverage a data-driven approach to learn the complex mapping between k-space and image domain, enabling efficient artifact removal, enhanced detail recovery, and high-quality reconstruction even at high acceleration factors. In this section, we give an overview of the widely used network models in the field of MRI reconstruction at first. Then, we delve into how these models are applied to both single-modality and multi-modality MRI data. In addition, we analyze MRI reconstruction methods that use different training strategies.

\subsection{Fundamental Models for MRI Reconstruction}

Through a detailed review of recent literature on deep learning-based MRI reconstruction, it is observed that CNNs, U-Nets, GANs, Transformers, diffusion-based architectures, deep unfolding networks, and Mamba are commonly used network architectures in this field. Therefore, it is necessary to provide a brief introduction to the fundamental concepts of these foundational models.

\subsubsection{CNNs} 

Convolutional Neural Networks (CNNs) have the key advantage of efficiently learning local features within images. CNNs utilize convolutional layers, pooling layers, and fully connected layers to progressively extract local features from input data and combine these features into more complex global information. Particularly when handling complex and noisy MRI data, CNNs can effectively extract useful information, thereby reconstructing high-quality images. Therefore, the application of convolutional neural networks in MRI reconstruction significantly enhances image quality, demonstrating strong robustness and adaptability.

\subsubsection{U-Nets} 

Initially, Ronneberger et al. introduced U-Net for image segmentation, achieving remarkable results and attracting significant attention from researchers~\cite{ronneberger2015u}. The U-Net architecture consists of a contracting path, a bottleneck layer, and an expanding path. The contracting path functions as an encoder, responsible for feature extraction, where the size of the input feature maps progressively decreases. The expanding path, on the other hand, acts as a decoder. The features obtained from the contracting path are passed through the bottleneck layer and then progressively expanded in the expanding path until they match the input size. 
U-Net is capable of capturing both high-level and low-level features, and it can efficiently extract features and perform multi-scale information fusion. MRI reconstruction can be viewed as an image transformation task, which is why the U-Net architecture is widely used in MRI reconstruction methods.

\subsubsection{Generative adversarial networks} 

Generative Adversarial Networks (GANs) were proposed by Ian Goodfellow et al. in 2014~\cite{gan}. GAN consists of two components: the generator and the discriminator. The generator aims to generate images as realistic as possible, while the task 
of the discriminator is to distinguish between real images and those generated by the generator. These two networks are optimized alternately through a competitive process. After multiple iterations, the generator learns to produce fake data that closely resembles real data, and the discriminator becomes increasingly difficult to distinguish between real and fake data. 

Cycle-GAN (Cycle-Consistent Generative Adversarial Network) is a variant of GAN, designed to address the problem of unsupervised image-to-image translation. The core idea is that when an image is transformed from the source domain to the target domain through the generator, and then mapped back to the source domain through the reverse generator, the original image should be recovered~\cite{zhu2017unpaired}. This process ensures that the mapping performed by the generator is meaningful, thereby preventing the generation of useless or inconsistent images. Cycle-GAN, by utilizing adversarial loss and cycle-consistency loss, can achieve high-quality image translation in an unsupervised manner. This is particularly useful for MRI image reconstruction tasks. In MRI reconstruction, Cycle-GAN effectively addresses the issue of data scarcity, especially when paired training data is unavailable, making it especially important for solving MRI reconstruction tasks with incomplete or hard-to-obtain datasets.

\subsubsection{Transformers}

Transformer is a deep learning architecture proposed by Vaswani et al. in 2017~\cite{vaswani2017attention}, initially designed for natural language processing (NLP) tasks. It is based on the self-attention mechanism~\cite{guo2022attention,zhang2023attention}, which effectively captures long-range dependencies in the data. With the introduction of Vision Transformer (ViT)~\cite{dosovitskiy2020image}, the Transformer architecture has successfully expanded into the computer vision domain, especially excelling in image processing. It divides the image into multiple small patches and then encodes these patches into a set of vectors, meeting the input requirements of the basic Transformer model. Through the multi-head self-attention mechanism, ViT can selectively focus on the most important regions of the image, effectively capturing global information. This enables ViT to better recover details and structural information in images. Swin Transformer~\cite{liu2021swin}, based on ViT, reduces computational complexity by computing self-attention within small windows and has demonstrated excellent performance across various computer vision tasks, showing promising prospects.

\subsubsection{Denoising diffusion probabilistic models} 

Denoising Diffusion Probabilistic Models (DDPMs) are a class of generative models initially proposed by Ho et al~\cite{ho2020denoising}. These models simulate a step-by-step noise-adding process, and then learn how to denoise in order to generate data. The training process of DDPMs typically involves two stages: the forward diffusion process and the reverse generation process. 

In the forward diffusion process(Eq.~\ref{eq:q1}), the model gradually adds a series of random Gaussian noise to the input image \( X_0 \), and after \( T \) steps, the noisy image \( X_T \) is obtained. The noisy image at the \( t \)-th step can be obtained by adding noise to the image at the \( t-1 \)-th step, as shown in Eq.~\ref{eq:q2}.

\begin{equation}
    q(x_t | x_{t-1}) = \mathcal{N}\left(x_t; \sqrt{1 - \beta_t} x_{t-1}, \beta_t I\right)
    \label{eq:q1}
\end{equation}

\begin{equation}
    x_t = \sqrt{1 - \beta_t} x_{t-1} + \epsilon \sqrt{\beta_t}; \quad \epsilon \sim \mathcal{N}(0, I)
    \label{eq:q2}
\end{equation}

Here, \( \mathcal{N} \) is the Gaussian distribution, \( \epsilon \) is the added noise, and \( I \) is the identity covariance matrix.

In the reverse generation process(Eq.~\ref{eq:reverse}), starting from a pure noise image, the target image is obtained through step-by-step denoising. Each step of the reverse denoising uses a neural network($p_\theta$) to fit the process, and trains by optimizing the variational lower bound on the simplified log-likelihood \cite{dayarathna2024deep}.

\begin{equation}
    p_\theta(x_{t-1} | x_t) = \mathcal{N}\left(x_{t-1}; \mu_\theta(x_t, t), \Sigma(x_t, t)\right)
    \label{eq:reverse}
\end{equation}

Typically, a trained diffusion model can denoise a noisy image to generate a sample close to the real image. For MRI reconstruction tasks, data consistency can be incorporated during the reverse diffusion phase. Specifically, the generated image sample is aligned with the k-space data of the real MRI image~\cite{wu2024adaptive,peng2022towards}, allowing for the generation of the high-quality MRI image we need.

\subsubsection{Deep unfolding networks} 

Deep Unfolding Networks (DUNs) are a class of network architectures that combine traditional optimization algorithms with deep learning models. The idea of unrolled networks was first introduced by Gregor et.al~\cite{gregor2010learning}, and then some key MRI reconstruction works have adopted Deep Unfolding Networks~\cite{gregor2010learning,hammernik2018learning,aggarwal2018modl,sun2016deep,schlemper2017deep}. Their foundational theory is based on optimization methods, particularly iterative optimization algorithms such as gradient descent and conjugate gradient methods. Deep unfolding networks "unfold" these optimization algorithms into a neural network structure, allowing each step of the traditional optimization process to be implemented by a layer of the neural network. Each layer in a deep unfolding network can be viewed as one iteration of the optimization algorithm. By simulating the optimization process with a deep network, the model can achieve the effect of multiple iterations of traditional optimization algorithms with fewer layers.

\subsubsection{Mamba} 

The Mamba model~\cite{gu2023mamba} is based on the State Space Model (SSM)~\cite{gu2021combining}, which can capture state representations and forecast subsequent states. The conversion from a one-dimensional function or sequence $x(t) \in \mathbb{R}$ to an output $y(t) \in \mathbb{R}$ is facilitated by a hidden state $h(t) \in \mathbb{R}^{\mathbb{N}}$. This process is generally realized using linear ordinary differential equations (ODEs), as described below.

\begin{equation}\label{eq01}
    h^{\prime}(t)=\mathbf{A} h(t)+\mathbf{B} x(t), \quad y(t)=\mathbf{C} h(t),
\end{equation}
where $\mathbf{A} \in \mathbb{R}^{\mathrm{N} \times \mathrm{N}}$ is the state matrix, and $\mathbf{B} \in \mathbb{R}^{\mathrm{N} \times 1}$ and $\mathbf{C} \in \mathbb{R}^{1 \times \mathrm{N}}$ are the projection parameters. To adapt ODEs for deep learning, the zero-order hold (ZOH) method is employed for discretization. By introducing a timescale parameter $\Delta$, the continuous-time matrices $\mathbf{A}$ and $\mathbf{B}$ are converted into their discrete counterparts, $\overline{\mathbf{A}}$ and $\overline{\mathbf{B}}$, through the following steps:

\begin{equation}\label{eq2}
\overline{\mathbf{A}}=\exp (\Delta \mathbf{A}), \quad \overline{\mathbf{B}}=(\Delta \mathbf{A})^{-1}(\exp (\Delta \mathbf{A})-\mathbf{I}) \cdot \Delta \mathbf{B} .
\end{equation}

Following discretization, Eq.\ref{eq01} is reformulated for discrete-time processing as:

\begin{equation}\label{eq3}
h_t=\overline{\mathbf{A}} h_{t-1}+\overline{\mathbf{B}} x_t, \quad y_t=\mathbf{C} h_t .
\end{equation}

Mamba introduces a selective mechanism, enabling the model to dynamically adjust the state transition matrix based on the input. Furthermore, Mamba significantly improves computational efficiency by optimizing GPU memory access and computation processes. Compared to the quadratic complexity of Transformers, the linear complexity of Mamba gives it a distinct advantage in handling long sequences.

\subsection{Applications of Single-modal MRI Reconstruction Methods}

Currently, deep learning-based MRI reconstruction methods can be broadly classified into single-modal reconstruction and multi-modal reconstruction. For clarity, this section first focus on the research progress of single-modal MRI reconstruction methods. Single-modal reconstruction methods typically rely on a single type of input data, and these methods are widely used in MRI image reconstruction, especially in reducing scan time and improving image quality. Table~\ref{tab:mri_methods} shows some representative MRI reconstruction methods. Here, Adv represents the adversarial loss, Cyc represents the cycle consistency loss, and Idt represents the identity loss. Next, we categorize and introduce these methods based on the underlying models they depend on. Table~\ref{tab:models} classifies the literature based on their underlying models.

\begin{table}[t]
    \centering
    \caption{Some representative MRI reconstruction methods.}
    \begin{tabular}{@{}cllllr@{}}
        \toprule
        Method & Anatomy & Architecture & 2D/3D & Loss & Reference \\ 
        \midrule
        SwinGAN & Brain & GAN & 2D & Adv/L2 & ~\cite{swingan} \\ 
        PDD-GAN & Brain & GAN & 2D & Adv/Cyc/Idt & ~\cite{li2024progressive} \\ 
        DSMENet & Brain/Knee & CNN & 2D & L2 & ~\cite{wang2023dsmenet} \\ 
        DLGAN & Brain/Knee & GAN & 2D & Adv & ~\cite{noor2024dlgan} \\ 
        RNLFNet & Brain/Knee & U-Net & 2D & L1 & ~\cite{zhou2023rnlfnet} \\ 
        CDF-Net & Brain/Knee & CNN & 2D & L1 & ~\cite{nitski2020cdf} \\ 
        ReconFormer & Brain/Knee & Transformer & 2D & L2 & ~\cite{guo2023reconformer} \\ 
        OCUCFormer & Brain/Knee/Cardiac & Transformer & 2D & L2 & ~\cite{al2024ocucformer} \\ 
        Nila-DC & Brain & Diffusion model & 2D & L2 & ~\cite{huang2024noise} \\ 
        C2E-DDPM &  Knee & Diffusion model & 2D & L2 & ~\cite{zhao2024center} \\ 
        MGDUN & Brain & Deep unfolding network & 2D & L1 & ~\cite{yang2023mgdun} \\ 
        The unrolled MCMR & Cardiac & Deep unfolding network & 2D & L1 & ~\cite{pan2024unrolled} \\ 
        MsFF-Net & Brain & CNN & 3D & L1 & ~\cite{kang20243d} \\
        MambaMIR &  Knee/Chest/Abdomen & Mamba & 2D & Adv/L2 & ~\cite{huang2024mambamir} \\
        NeSVoR & Brain & MLP & 3D &  L2 hybrid loss function & ~\cite{xu2023nesvor} \\
       
        \bottomrule
    \end{tabular}
    \label{tab:mri_methods}
\end{table}

\begin{table}[t]
    \centering
    \caption{This table classifies MRI reconstruction methods according to their foundational models. Each row represents a different backbone architecture used in MRI reconstruction, along with a list of references that correspond to the use of that architecture in the literature.}
    \begin{tabular}{@{}cr@{}}
        \toprule
        Backbone & Reference  \\ 
        \midrule
        CNN & \cite{zou2022joint,guo2021over,chen2022pyramid,wang2024crnn,yiasemis2022recurrent,sun2025fourier,bongratz2024neural,chen2022accelerating,wang2023dsmenet,ren2022complex,liu2022undersampled,mittal20243d,ding2022mri,liu2021universal,kang20243d,corona2021variational,wang2023mhan} \\ 
        U-Net & ~\cite{nitski2020cdf,zhou2023rnlfnet,wang2024dpfnet,yang2024attention,huang2021dynamic,chen2023surfflow,li2024radial} \\ 
        GAN & ~\cite{chen2021wavelet,lv2021pic,yaqub2022gan,han2021madgan,yuan2020sara,cole2020unsupervised,li2023cs,zhang20213d,murugesan2019recon,li2021modified,liu2022dbgan,yang2017dagan,deora2020structure,swingan,noor2024dlgan,li2024progressive,sood20213d,elmas2022federated,zhou2021efficient,wu2024compressed,lv2021transfer,sangeetha2024c2} \\ 
        Transformer & ~\cite{huang2022swin,huang2022fast,korkmaz2021deep,zhao2024diffgan,hu2022trans,shen2024magnetic,xu2023learning,du2023transformer,alghallabi2023accelerated,lin2022vision,korkmaz2022mri,ekanayake2022multi,feng2022multimodal,zhao2022k,wu2023deep,sheng2024cascade,guo2023reconformer,yi2023frequency,feng2021task,gao2022projection,pan2023global,feng2023learning,ekanayake2024mcstra,lyu2023region,al2024ocucformer} \\ 
        Denoising diffusion model & ~\cite{webber2024diffusion,luo2023bayesian,bian2024diffusion,chung2022score,levac2023mri,kazerouni2023diffusion,safari2024adaptive,jiang2024fast,geng2024dp,huang2024noise,zhao2024center,gungor2023adaptive,korkmaz2023self,ozturkler2023smrd,tan2024fetal,guan2024correlated} \\ 
        Deep unfolding network & ~\cite{luo2025deep,luo2024joint,lei2023deep,zhang2025re,ma2024attention,wang2023deep,zhang2024t2lr,gunel2022scale,aghabiglou2022deep,wang2021memory,fan2023interpretable,yang2023mgdun,pan2024unrolled} \\ 
       Mamba & ~\cite{gu2023mamba,korkmaz2024mambarecon,huang2024mambamir,huang2025enhancing} \\ 
        \bottomrule
    \end{tabular}
    \label{tab:models}
\end{table}

\subsubsection{CNNs} 

CNN is widely used in the field of MRI reconstruction.
In 2021, Liu et al. first proposed a universal network for MRI reconstruction~\cite{liu2021universal}, which is based on CNN. After training, this network requires minimal additional training to achieve good reconstruction performance for new anatomical structures. The introduction of this method effectively addresses the following issue: Many MRI reconstruction methods are designed for specific anatomical structures. After model training, their ability to generalize to other anatomical structures is limited.

In 2022, Chen et al. proposed a novel framework, namely PU-Network-PS (UNS), to accelerate Deeply Cascaded Convolutional Networks (DCCNs) for MRI reconstruction~\cite{chen2022accelerating}. The approach primarily enhances computational efficiency by shifting computations to a lower-resolution space. Experimental results show that this method can improve the reconstruction performance of the DCCNs model. Unlike previous reconstruction methods, Ren et al. proposed a new complex-valued dual-domain dilated convolution neural network (C3DNet)~\cite{ren2022complex}. This method simultaneously extracts complex-valued features from both k-space and image domain data by using complex-valued convolution. Liu et al. proposed a Side Information-Guided Normalization (SIGN) module~\cite{liu2022undersampled}, which enhances the reconstruction quality by using side information from MRI as normalization parameters within the convolutional network. Ding et al. proposed to directly complete the missing and corrupted k-space data using a specially designed interpolation deep neural network~\cite{ding2022mri}, which incorporates convolutional layers to regularize the data. Notably, the network they proposed offers good interpretability and also demonstrates impressive computational efficiency.

In 2023, to tackle the insufficient attention to different components in zero-filled (ZF) MR images, Wang et al. proposed a Detail and Structure Mutually Enhancing Network (DSMENet)~\cite{wang2023dsmenet}. The Structure Reconstruction UNet (SRUN) and the Detail Feature Refinement Module (DFRM) are used to learn structural information and image detail information, respectively.

\subsubsection{U-Nets} 

Many studies have adopted the U-Net architecture for MRI reconstruction~\cite{nitski2020cdf,zhou2023rnlfnet,wang2024dpfnet,zhao2024center,swingan,li2024progressive,zhou2020dudornet,eo2018kiki,ran2020md,liu2023diik,sun2020dual,liu2022dual,wang2024dct,wang2019accelerated,zhang2021dual,zhang2025fdudoclnet,seo2021dual,lu2025dffki}. They incorporate novel mechanisms to enhance model performance and generate higher-quality reconstructed images.

In 2020, Nitski et al. proposed a cross-domain fusion network (CDF-Net)~\cite{nitski2020cdf}, which consists of three components: a k-space reconstruction module, an image domain reconstruction module, and a module for fusing k-space and image domain information. All three modules are based on the U-Net architecture, and the CDF-Net utilizes both the image domain and k-space domain in an end-to-end manner.

In 2021, Huang et al. proposed a novel dynamic MRI reconstruction approach called MODRN and an end-to-end improved version called MODRN(e2e)~\cite{huang2021dynamic}. Its motion estimation module, based on the U-Net architecture, can effectively correct images according to the motion field. Experimental results have demonstrated that this method outperforms several of the most advanced methods at the time. This method is specifically designed for cardiac MRI reconstruction. It makes full use of motion information to reduce motion artifacts. Cardiac MRI reconstruction is crucial for doctors to diagnose diseases related to the heart, and there are also many studies specifically aimed at cardiac MRI reconstruction~\cite{huang2021dynamic,groun2022novel,lyu2023region,pan2024unrolled,basit2023accelerating}. 

\begin{figure}[t]  
    \centering
    \includegraphics[width=0.5\textwidth]{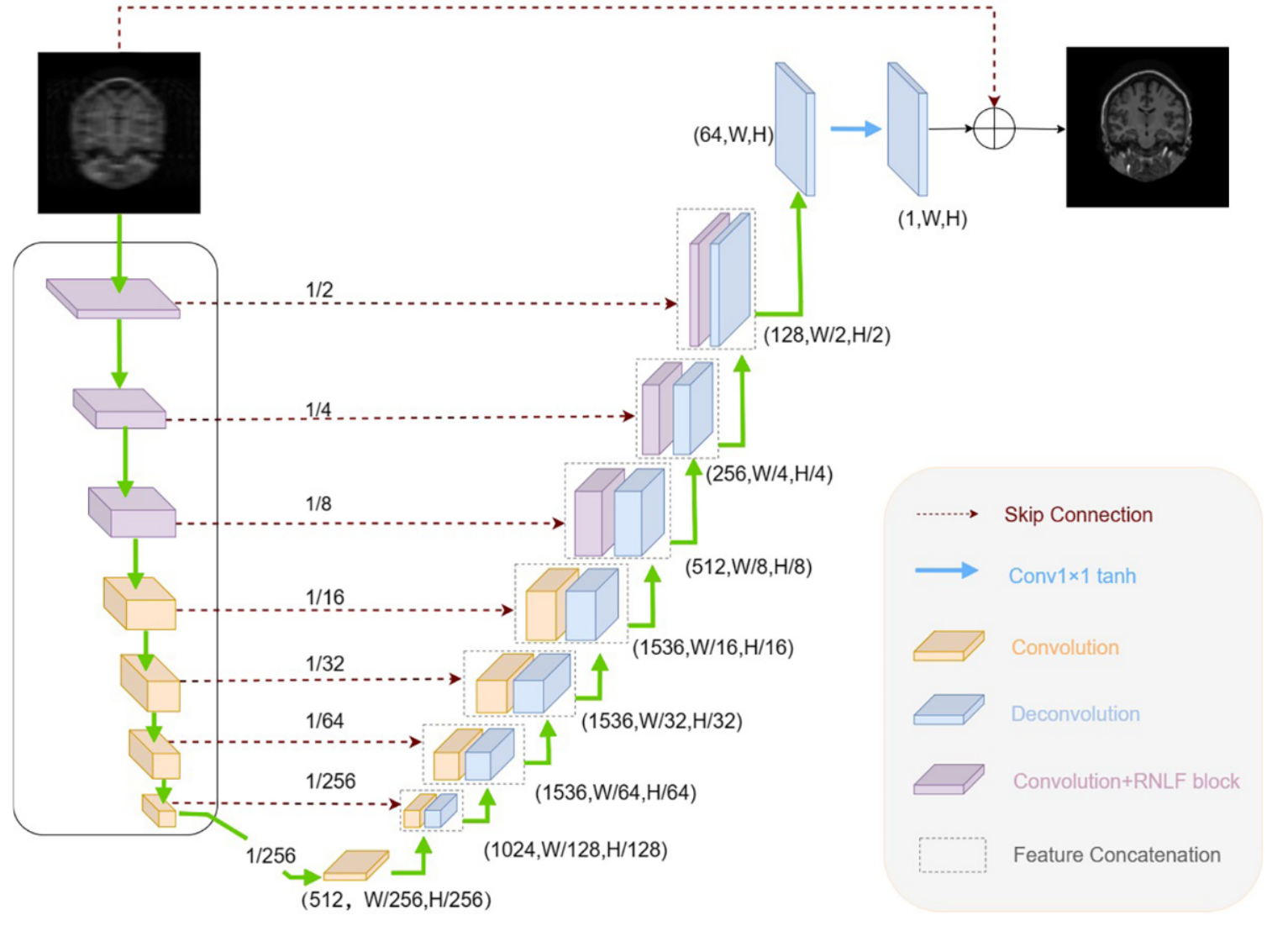}  
    \caption{The architecture diagram of RNLFNet~\cite{zhou2023rnlfnet}, a typical U-Net reconstruction network.}  
    \label{fig:rnlf}  
\end{figure}

In 2023, Zhou et al. proposed a new deep Residual NonLocal Fourier Network (RNLFNet)~\cite{zhou2023rnlfnet}, which simultaneously learns information from both the image domain and k-space domain. The architecture diagram of RNLFNet is shown in Figure~\ref{fig:rnlf}. Using U-Net as the backbone framework and combining Fourier transforms with a self-attention mechanism, the network effectively captures long-range information in the frequency domain, significantly improving the reconstruction quality. Chen et al. proposed a method called SurfFlow~\cite{chen2023surfflow}, specifically designed for rapid and accurate cortical surface reconstruction from infant brain MRI, addressing the limitations of previous methods that were tailored for adult brain MRI reconstruction. The network consists of three deformation blocks, each based on a 3D U-Net architecture.

In 2024, Li et al. proposed a new complex-valued convolutional neural network, namely Dense-U-Dense Net (DUD-Net)~\cite{li2024radial}. It fully utilizes the relationship between the real and imaginary parts of k-space data. By using U-Net to perform interpolation estimation on k-space data, the network effectively improved the reconstruction quality. To accelerate the reconstruction of chemical exchange saturation transfer (CEST) magnetic resonance imaging (MRI), Yang et al. proposed an attention-based multi-offset deep learning reconstruction network (AMO-CEST)~\cite{yang2024attention}. The network also uses the U-Net framework combined with residual modules to extract multi-scale features. Many deep learning methods are based on parallel imaging, aiming to improve the fusion of multi-coil data to enhance reconstruction quality~\cite{arvinte2021deep,gan2021ss,sun2023joint,hashemizadehkolowri2021simultaneous,murugesan2021deep,liu2023high}. For example, Wang et al. proposed a new Dual-domain Parallel Fusion Reconstruction Network (DPFNet)~\cite{wang2024dpfnet}, which uses U-Net as the backbone network to address the issue of insufficient reconstruction details in multi-coil MRI reconstruction tasks.

\subsubsection{GANs} 

Generative adversarial networks, due to their generative nature, are particularly well-suited for image reconstruction tasks.

In 2017, Yang et al. proposed a conditional Generative Adversarial Networks-based model (DAGAN)~\cite{yang2017dagan}. This research is one of the more representative early works that applied GANs to MRI reconstruction. Notably, in this study, they introduced a refinement learning method, similar to the concept of ResNet, which allows the generator to learn only the missing information compared to fully sampled images. This approach facilitates training and reduces inference time, making it suitable for real-time processing.

In 2020,  Deora et al. proposed a GAN-based method that uses a patch-based discriminator, which focuses more on preserving high-frequency information~\cite{deora2020structure}. This is reflected in the retention of more fine texture details. The method reduces reconstruction time and achieves better reconstruction quality compared to the state-of-the-art methods at that time. It is worth noting that it also reduces the inference time for real-time reconstruction.

In 2021, Sood et al. proposed a novel multi-image super-resolution generative adversarial network (miSRGAN)~\cite{sood20213d}, which can reconstruct 3D MRI images and also learn how to perform 3D registration during the reconstruction process. To enhance the reconstruction quality in highly undersampled spaces, Zhou et al. proposed an efficient structurally-strengthened Generative Adversarial Network~\cite{zhou2021efficient}. Given the difficulty in obtaining a large amount of patient k-space data, Lv et al. considered using transfer learning and proposed a novel approach~\cite{lv2021transfer}. By combining parallel imaging with the GAN model (PI-GAN), they were able to transfer a model pre-trained on a large dataset (e.g., the Calgary-Campinas brain dataset) to small-scale datasets in the target domain (e.g., knee, liver, and brain tumor MRI data). This enhanced the generalization ability of these small-sample networks.

In 2022, Elmas et al. proposed Federated Learning of Generative Image Priors (FedGIMP) for MRI reconstruction~\cite{elmas2022federated}. Through Federated Learning (FL) and using GAN as the framework, they achieved data sharing across multiple institutions. It facilitates training using data from multiple institutions while ensuring privacy.

\begin{figure}[t]  
    \centering
    \includegraphics[width=0.6\textwidth]{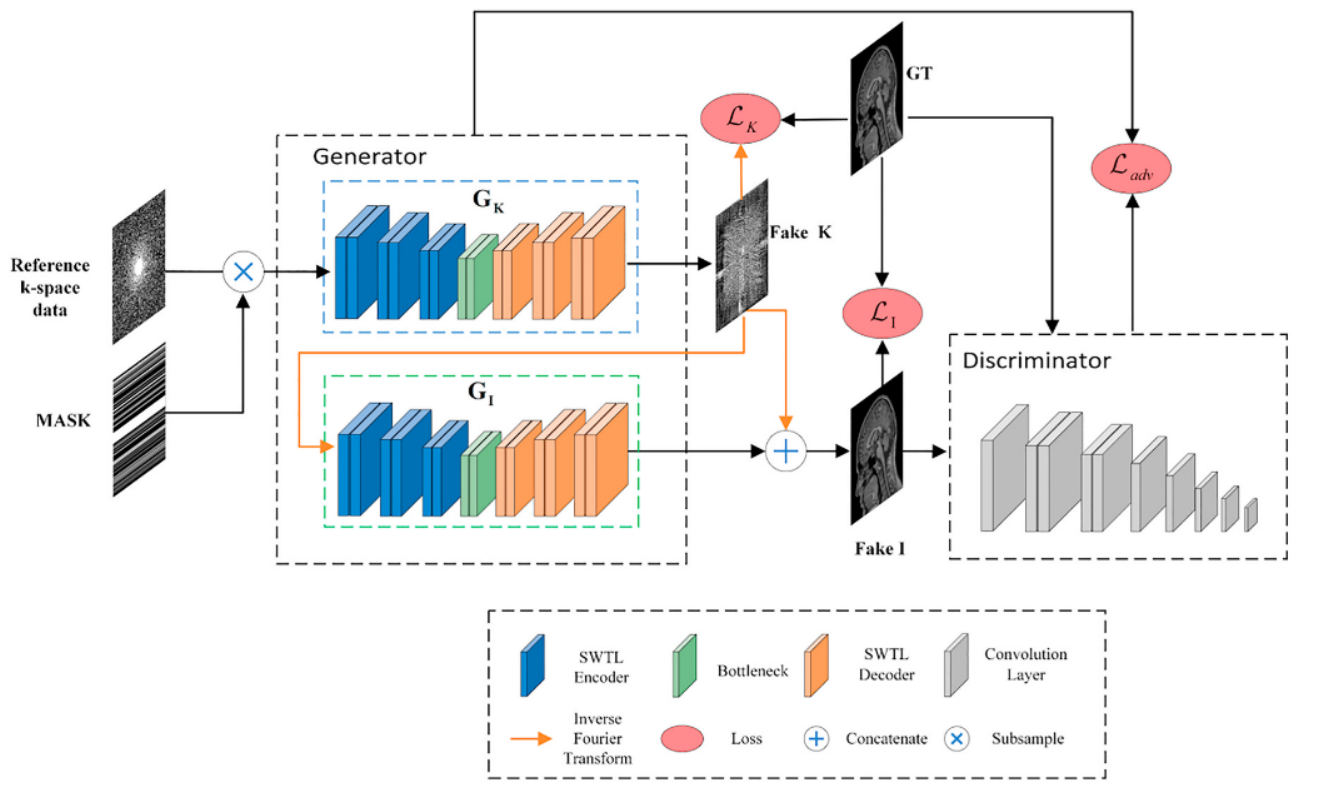}  
    \caption{The architecture diagram of SwinGAN~\cite{swingan}.}  
    \label{fig:swingan}  
\end{figure}

In 2023, Zhao et al. proposed SwinGAN~\cite{swingan}. It is a representative method in recent years that uses GAN for MRI reconstruction. It employs GAN as the backbone network and integrates the Swin Transformer to effectively capture long-range dependencies. The architecture diagram is shown in Figure~\ref{fig:swingan}. In this method, \( G_k \) and \( G_I \) are the generators in the k-space domain and image domain, respectively. The method simultaneously utilizes information from both the k-space and image domains for MRI reconstruction. Experimental results demonstrate that this method effectively utilizes both local and global information, significantly improving the reconstruction quality.

In 2024, Noor et al. proposed a novel DLGAN (Deep Learning Generative Adversarial Network) model~\cite{noor2024dlgan}, specifically designed for knee and brain image reconstruction. It effectively improves reconstruction quality and reduces artifacts. To improve the generalization ability of reconstruction in highly undersampled spaces, Wu et al. proposed DAPGAN(a deep adaptive perceptual generative adversarial network)~\cite{wu2024compressed}. This method introduces a novel perceptual feature guidance (PFG) mechanism, which helps capture features at various scales that are beneficial for reconstruction. Experimental results demonstrate that it effectively reconstructs detailed parts of the image.  Sangeetha et al. proposed a novel deep learning-based \( C^{2} \) LOG-GAN for super-resolution MRI reconstruction~\cite{sangeetha2024c2}. This method utilizes two GAN structures to achieve better reconstruction results. The first GAN generates the low-high-resolution image from the Low-resolution (LR) image. Then, the second GAN generates the Super-resolution (SR) image from the high-resolution image.

\subsubsection{Transformers} 

The Transformer, due to its outstanding ability to capture long-range dependencies, is widely used in the field of MRI reconstruction. 

In 2022, Gao et al. proposed a novel data augmentation method that can enhance the performance in reconstruction tasks involving non-Cartesian trajectories such as the radial trajectory~\cite{gao2022projection}. 

\begin{figure}[t]  
    \centering
    \includegraphics[width=0.8\textwidth]{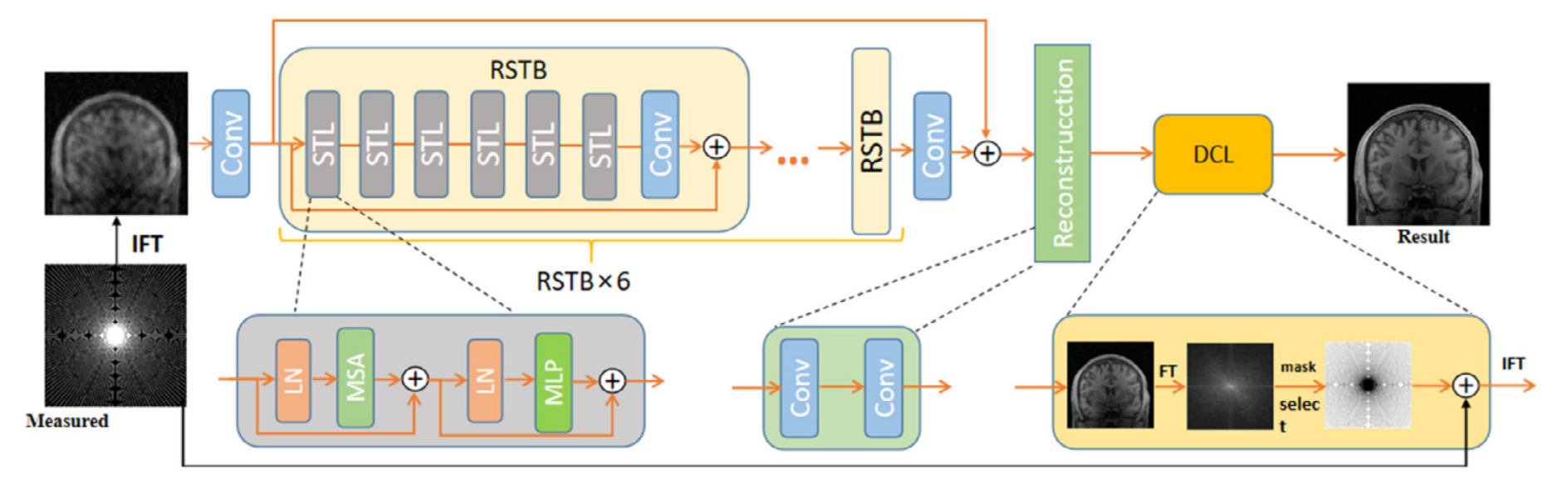}  
    \caption{A typical architecture for MRI reconstruction using Transformer, RSTB denotes Residual Swin Transformer Block and DCL denotes data consistency layer. From~\cite{wu2023deep}.}  
    \label{fig:transformerxx}  
\end{figure}

In 2023, Wu et al. utilized the Swin Transformer as the backbone network for MRI reconstruction and further enhanced the reconstruction quality by incorporating data consistency~\cite{wu2023deep}. They demonstrated that using the Swin Transformer for MRI reconstruction can more effectively extract deep image features. The architecture of the method is shown in Figure~\ref{fig:transformerxx}. Guo et al. proposed a recurrent Transformer model, namely ReconFormer~\cite{guo2023reconformer}, which designs a recursive attention mechanism that can utilize multi-scale information to improve reconstruction quality. Compared to other models, it has a small number of parameters, making it easy to train. Similarly, to improve computational efficiency while fully considering frequency information and non-local similarity, Yi et al. proposed Frequency Learning via Multi-scale Fourier Transformer for MRI Reconstruction (FMTNet)~\cite{yi2023frequency}. This method separately employs high-frequency and low-frequency branches to learn relevant frequency details. Moreover, it uses Fourier convolution to replace the commonly used self-attention mechanism, allowing for more efficient learning of long-range dependencies with fewer computational resources. Pan et al. proposed a novel Transformer-based k-space Global Interpolation Network, termed k-GIN~\cite{pan2023global}. This network first interpolates the undersampled k-space data and then transforms it to the image domain using the Fourier transform. By learning the global dependencies in the frequency information, the network can effectively perform interpolation. To enable multiple institutions to share data for MRI reconstruction without compromising privacy, Feng et al. proposed a new federated learning algorithm called FedPR~\cite{feng2023learning}. This algorithm significantly reduces the communication cost associated with federated learning in MRI reconstruction tasks and improves computational efficiency. To fully capture the spatiotemporal information in cardiac cine MRI reconstruction, Lyu et al. proposed a region-focused multi-view Transformer-based generative adversarial network for cardiac cine MRI reconstruction~\cite{lyu2023region}. For this task, the method achieved the best performance at that time.

In 2024, Fahim et al. proposed an Over-Complete Under-Complete Transformer network (OCUCFormer)~\cite{al2024ocucformer}. This method not only promotes MRI reconstruction for both single-coil and multi-coil scenarios but also employs a self-supervised training approach, reducing the reliance on paired data. To fully leverage the physics of MRI, Ekanayake et al. proposed a novel physics-based Transformer model titled the Multi-branch Cascaded Swin Transformers (McSTRA)~\cite{ekanayake2024mcstra}. By integrating MRI-related physical concepts with the Swin Transformer, this model improves reconstruction quality and robustness. Previous methods largely relied excessively on either CNNs or Swin Transformer blocks. The former often led to a limited receptive field, making it difficult to capture global features, while the latter resulted in an overly large receptive field, increasing model complexity. To address this issue, Sheng et al. proposed a cascade dual-domain Swin-Conv Unet for reconstruction (CDSCU-Net)~\cite{sheng2024cascade}, which combines the strengths of CNNs and Swin Transformer. This model is capable of capturing both local and global features simultaneously, thereby effectively improving reconstruction quality. 

\subsubsection{Denoising diffusion models} 

In recent years, with the continuous development of diffusion models in the field of image generation, more and more MRI reconstruction methods have also begun to employ diffusion models.

\begin{figure}[t]  
    \centering
    \includegraphics[width=0.5\textwidth]{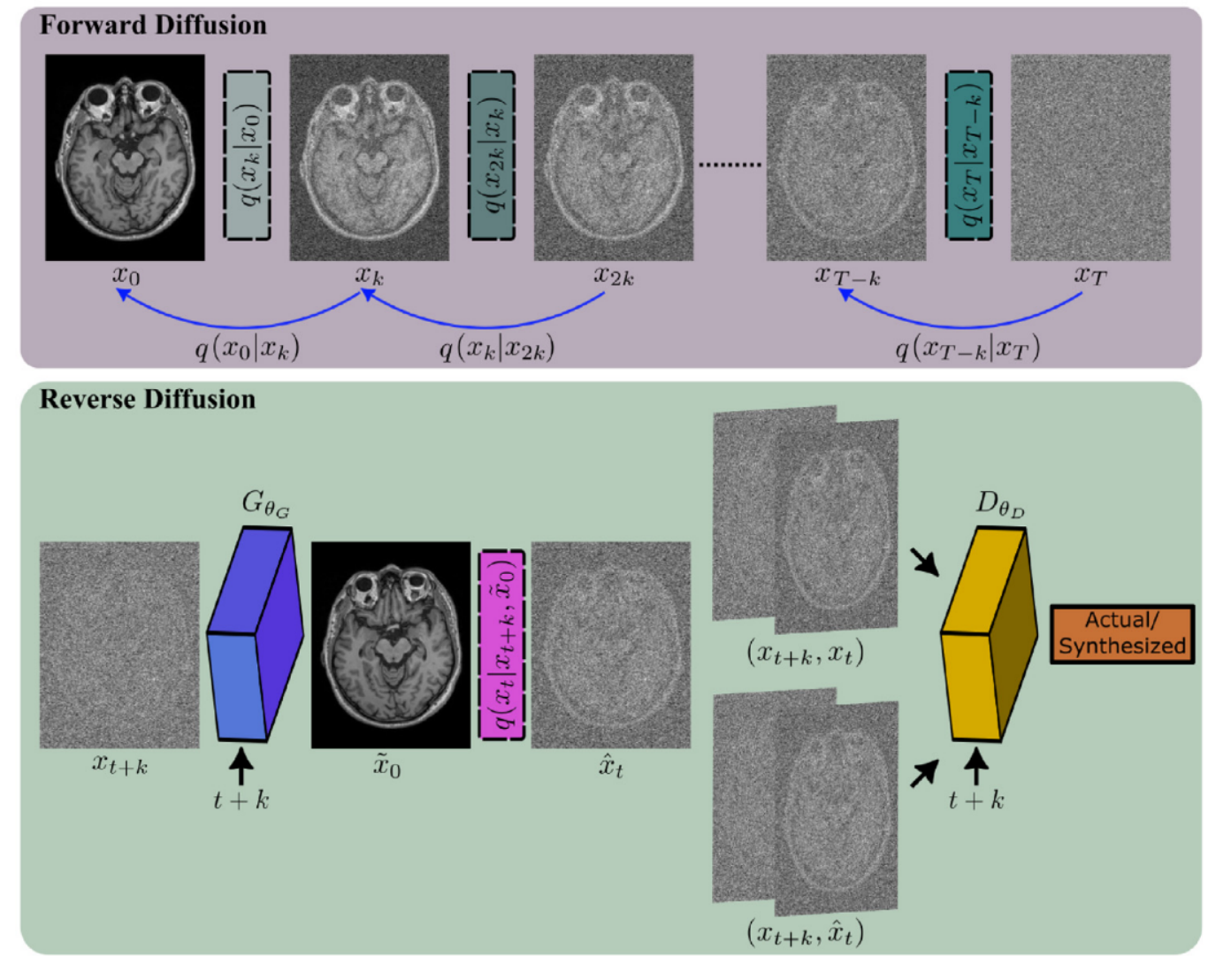}  
    \caption{The architecture diagram of AdaDiff~\cite{gungor2023adaptive}.}  
    \label{fig:adadiff}  
\end{figure}

In 2023, Güngör et al. proposed the first adaptive diffusion prior for MRI reconstruction, AdaDiff~\cite{gungor2023adaptive}. The architecture of AdaDiff is shown in Figure~\ref{fig:adadiff}. During the forward diffusion process, noise is incrementally added to the original image until it becomes a pure noise image. In the reverse diffusion process, the pure noise image is fed into the generator to produce a reconstructed image, which is then combined with a reference image and input into the discriminator for training. This method divides the reconstruction process into two stages. In the first stage, a rapid diffusion process is conducted to obtain an initial reconstructed image. In the second stage, the prior is updated to further improve the quality of the initial reconstructed image. Experiments have shown that the AdaDiff method achieves excellent results. Yilmaz Korkmaz et al. proposed Self-Supervised Diffusion Reconstruction (SSDiffRecon)~\cite{korkmaz2023self}. This method represents the conditional diffusion process in traditional diffusion models as an unrolled architecture. This architecture leverages a combination of cross-attention mechanisms and data consistency to better reconstruct the missing information in k-space. Notably, this approach also achieves unsupervised training, reducing the dependence on paired data. Experiments have shown that SSDiffRecon outperforms previous methods, whether in supervised or unsupervised settings. Due to the sensitivity of diffusion models to distribution shifts during inference and the need for careful tuning of hyperparameters on the validation set when using diffusion models, a solution has been proposed in the form of SURE-based MRI Reconstruction with Diffusion models (SMRD)~\cite{ozturkler2023smrd}. This method can estimate reconstruction error during testing and automatically adjust the inference hyperparameters without the need for validation tuning. SMRD is the first approach to incorporate Stein’s Unbiased Risk Estimator (SURE) into the sampling stage of diffusion models for automatic hyperparameter selection. The reconstruction quality of SMRD also outperforms prior diffusion model-based methods.

In 2024, Zhao et al. proposed a novel center-to-edge Denoising Diffusion Probabilistic Model (C2E-DDPM)~\cite{zhao2024center}, which is the first to explore the impact of diffusion models on dual-domain information. This method incorporates a module that integrates information from both the frequency and image domains, effectively fusing the data from these two domains. Experiments demonstrated that this approach achieved the best performance on the FastMRI dataset at the time. To achieve MRI reconstruction in highly under-sampled spaces while preserving more fine details, Guan et al. proposed an innovative principle named the Correlated and Multi-frequency Diffusion Model (CM-DM)~\cite{guan2024correlated}. Notably, the fundamental principle of this method lies in effectively combining or replacing components to optimize performance. This approach enables the diffusion process to converge more rapidly and significantly improves reconstruction quality. There are some studies specifically for fetal MRI reconstruction~\cite{tan2024fetal,cordero2022fetal}, Tan et al. proposed a novel coarse-to-fine self-supervised fetal brain MRI Radiation Diffusion Generation Model (RDGM)~\cite{tan2024fetal}. This method utilizes the forward diffusion process in diffusion models to transform the global intensity discriminant information in the 3D volume, effectively improving reconstruction quality. Typically, when diffusion models are applied to MRI reconstruction, they first gradually add artificial noise during the forward diffusion process and then iteratively remove this artificial noise while imposing data consistency to reconstruct the image. Ideally, this approach can achieve satisfactory reconstruction quality. However, in practice, MRI images already contain inherent noise during the acquisition process, which can interfere with the model's noise prediction during the denoising process, leading to suboptimal reconstruction quality. To address this issue, Huang et al. proposed a posterior sampling strategy with a novel Noise Level Adaptive Data Consistency (Nila-DC) operation~\cite{huang2024noise}. This method analyzes the input data to estimate the inherent noise level, thereby automatically adapting to the actual noise level and improving reconstruction quality.

\subsubsection{Deep unfolding networks} 

\begin{figure}[t]  
    \centering
    \includegraphics[width=0.8\textwidth]{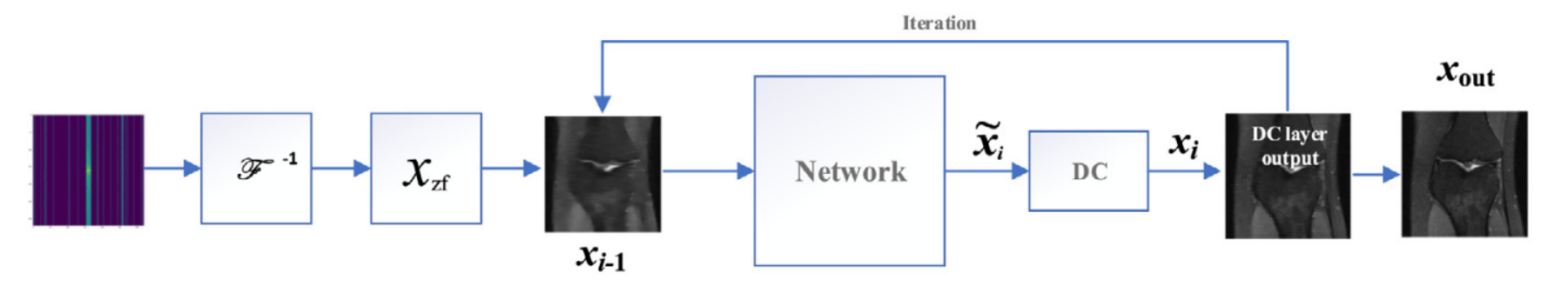}  
    \caption{A typical framework of an MRI reconstruction method based on deep unfolding networks, from~\cite{aghabiglou2022deep}.}  
    \label{fig:unroll}  
\end{figure}

With the development of deep unfolding networks in the field of image processing, a large number of MRI reconstruction methods have adopted deep unfolding networks. Deep unfolding network-based methods can provide advanced reconstruction performance and fast inference time for MRI reconstruction tasks~\cite{gunel2022scale}. A typical framework of an MRI reconstruction method based on deep unfolding networks is shown in Figure~\ref{fig:unroll}. The k-space data is first subjected to an inverse Fourier transform to obtain the zero-filled image (\( X_{zf} \)). Subsequently, it undergoes several iterations, with each iteration passing through a data consistency layer to enhance the reconstruction quality. The final reconstructed image is obtained at the end of this process.

In 2022, Aghabiglou et al. introduced an adaptive noise level parameter to the unfolding structure, inspired by conventional iterative thresholding-based reconstruction models~\cite{aghabiglou2022deep}. By adding this new noise level parameter as an additional input to the network, it acts as an evolving regularizer for the image manipulation strength of the network over the unfolding iterations. This approach effectively converges to better reconstruction results compared to state-of-the-art methods without such an adaptive parameter. Gunel et al. proposed modeling the proximal operators of unrolled neural networks with scale-equivariant convolutional neural networks~\cite{gunel2022scale}. This method can fully utilize the limited data for training.

In 2024, Pan et al. proposed a learning-based and unrolled MCMR framework~\cite{pan2024unrolled} for cardiac MR imaging reconstruction. It can achieve accurate Motion-Compensated MR reconstruction (MCMR). This method simultaneously addresses motion estimation and the reconstruction task, obtaining good reconstruction performance through joint optimization of these two tasks. 

\subsubsection{Mambas} 

\begin{figure}[t]  
    \centering
    \includegraphics[width=0.8\textwidth]{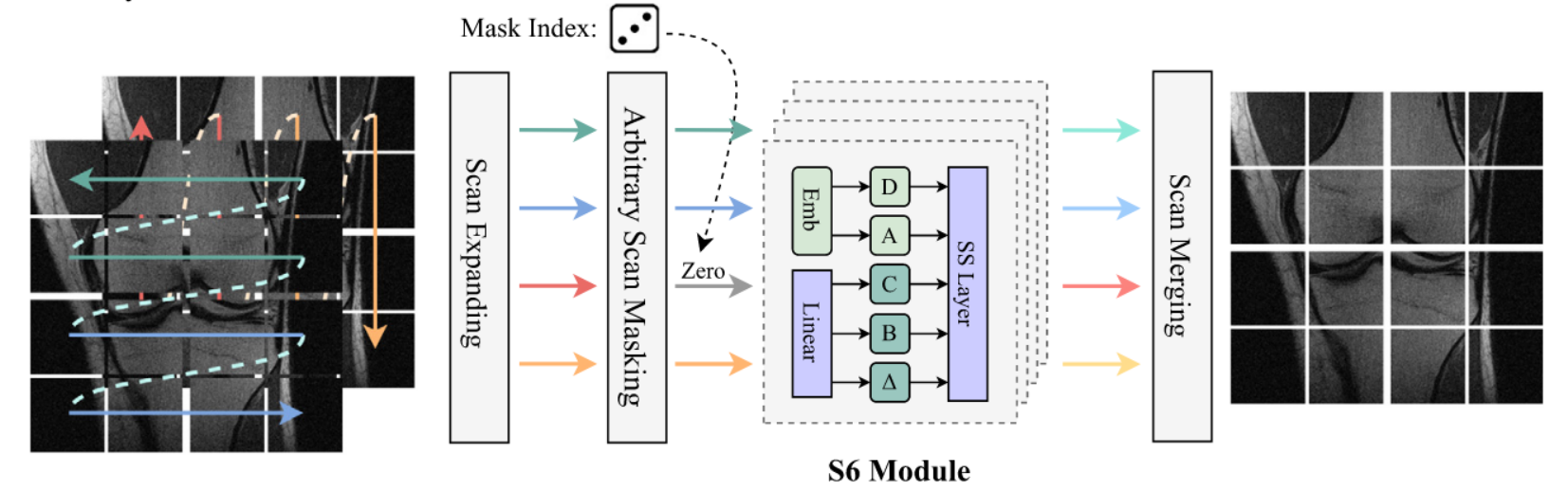}  
    \caption{The framework of MambaMIR~\cite{huang2025enhancing}.}  
    \label{fig:mamba}  
\end{figure}

The Mamba model~\cite{gu2023mamba} has garnered significant attention for its powerful capabilities in handling long sequences. Vision Mamba is now widely applied to computer vision tasks, such as image classification~\cite{he2024igroupss,li2024mambahsi},segmentation~\cite{yang2024remamber,xing2024segmamba}, and super-resolution~\cite{ji2024deform,ji2025generation,ji2025self}.
Recently, several methods~\cite{korkmaz2024mambarecon,huang2024mambamir,huang2025enhancing,zou2024mmr} applying it to MRI reconstruction have been proposed. 
Since Mamba can achieve similar functionality to the Transformer with higher computational efficiency, it is a very logical idea to consider using Mamba as a substitute for the Transformer. Korkmaz et al. proposed a new MRI reconstruction method based on Mamba~\cite{korkmaz2024mambarecon}, and experiments demonstrated that this method achieved the best performance at that time.
Huang et al. proposed a Mamba-based medical image reconstruction model named MambaMIR, as well as its GAN-based variant, MambaMIR-GAN~\cite{huang2024mambamir}. Notably, this method is not limited to MRI reconstruction but can also be applied to other types of medical image reconstruction. Additionally, it offers a certain degree of interpretability. Subsequently, Huang et al. extended their work by incorporating wavelet transforms, which further enhanced perceptual quality~\cite{huang2025enhancing}. The framework of MambaMIR is shown in Figure~\ref{fig:mamba}. MambaMIR is a general framework that can be used for medical image reconstruction and uncertainty estimation across different imaging modalities. Experiments conducted on several representative datasets demonstrated that MambaMIR achieved the best reconstruction fidelity, while MambaMIR-GAN exhibited the best perceptual quality. 

\subsection{Multi-modal MRI Reconstruction}

In the field of MRI reconstruction, multi-modal approaches have become a very popular direction, and many significant advancements have been made in this area~\cite{feng2021deep,li2023deep,wang2024spatial,luo2023effective,chen2024multi,chen2024fefa,xuan2022multimodal,casamitjana2022robust,wei2024misalignment,bian2022learnable,wang2024spatial,wei2022undersampled,yan2024cross}. Multi-contrast MRI Reconstruction involves combining MRI images from different weighted modalities (such as T1-weighted, T2-weighted, etc.) to generate high-quality images. For example, T1-weighted images provide clear anatomical structures, while T2-weighted images can highlight areas of edema. Different weighting methods describe the same anatomical region and provide complementary information, helping clinicians to analyze disease features more comprehensively. Therefore, the goal of multi-modal MRI reconstruction is to fuse images from different weighted modalities to compensate for the limitations of single-modality images and to enhance the overall image quality. Since different modality images may have positional discrepancies when describing the same anatomical region, they are typically aligned through image registration, after which algorithms are used for image fusion. This process results in a final image that integrates the information from all the weighted modalities. 

In multi-modal MRI reconstruction, it is common to provide one or more fully sampled reference modalities (such as T2) as auxiliary information for the reconstruction of an undersampled target modality (such as T1). However, this may raise two questions:

\textit{(1) The reason for MRI reconstruction is that MRI scanning time is long and requires undersampling. However, when performing multi-modal reconstruction, why is it still necessary to provide one or more fully sampled reference modalities (such as T2) as auxiliary information? Isn't it a waste of time to acquire an additional fully sampled reference modality?}

\textbf{Answer:} The acquisition of fully sampled auxiliary modalities is not a waste of time. First, the fully sampled reference modality can provide more prior information to help reconstruct the target modality, leading to a higher-quality reconstructed image. Second, the fully sampled reference modality itself contains unique information that can add additional value to the reconstruction process. It is also worth noting that in multi-modal MRI reconstruction, modalities acquired at different time points can be used as auxiliary information. Specifically, some modalities (such as T1, T2, etc.) do not undergo significant changes over time, so previously acquired fully sampled modalities can be used as references for current image reconstruction. This way, when changes over time are minimal, there is no need to re-acquire the fully sampled auxiliary modality, thus saving scan time.

\textit{(2) Since there is already a fully sampled reference mode, why does the doctor not use this fully sampled modality directly for diagnosis, but still reconstruct the undersampled modality?}

\textbf{Answer:} Although different weighted modalities share some common information, each modality also has its unique imaging characteristics and information. Even if we have a fully sampled reference modality (such as T2), it cannot completely replace the undersampled target modality (such as T1). Therefore, even with a fully sampled reference modality, reconstructing the undersampled target modality remains important. This is because the biological information revealed by each modality is complementary, and directly using the fully sampled modality does not fully exploit the potential of the undersampled modality, especially when precise lesion detection and tissue structure analysis are required.

 The introduction of multi-modal reconstruction is primarily based on the specific problems these methods address or the innovative techniques they employ.

It can be thought that the use of multi-modal reconstruction is mainly to make full use of redundant information between different modalities to guide the target modal reconstruction, and the purpose is to improve the reconstruction quality~\cite{feng2021deep,li2023deep,wang2024spatial,luo2023effective,chen2024multi}. When most MRI reconstruction methods focused solely on single-modal reconstruction, Feng et al. proposed a Multi-modal Aggregation Network for MRI Image Reconstruction with Auxiliary Modality (MARIO)~\cite{feng2021deep}. By aggregating information from multiple modalities, they achieved more comprehensive feature fusion and effectively improved the quality of reconstruction. Their experiments also demonstrated that their method outperformed the state-of-the-art reconstruction methods at the time in terms of artifact reduction. When the MRI reconstruction field began to focus on multi-modal reconstruction, although the use of fully sampled reference modalities helped improve reconstruction quality, it was still affected by aliasing artifacts. To further enhance reconstruction quality, Li et al. proposed a k-Space Partition-based Convolutional Network (kSPCN)~\cite{li2023deep}. By simultaneously partitioning the k-space of the reference modality and under-sampled modality into several subregions with gradually increasing sizes, and then reconstructing one by one according to subregions, their method effectively improved the reconstruction quality. Similarly, to enhance the performance of deep unfolding networks in MRI reconstruction, Chen et al. proposed a novel information-growth holistic unfolding network, named IHUN~\cite{chen2024multi}, which is capable of aggregating high-frequency information from different modalities to recover fine details.

Other motivations, such as addressing misalignment issues, ultimately also aim to improve reconstruction quality. Spatial misalignment is an inevitable issue when dealing with multimodal methods, and many current approaches are exploring ways to address this problem~\cite{chen2024fefa,xuan2022multimodal,casamitjana2022robust,wei2024misalignment}. To address this, Xuan et al. proposed a spatial alignment network to compensate for this spatial misalignment and improve the quality of multi-modal reconstruction~\cite{xuan2022multimodal}. Additionally, they designed a cross-modal loss combined with the reconstruction loss, enabling the spatial alignment network and reconstruction network to be trained simultaneously. Similarly, to address the multi-modal data alignment issue, Adrià Casamitjana et al. proposed a probabilistic model of spatial deformation~\cite{casamitjana2022robust}. Importantly, they also provided a registration of the reconstructed volume to MNI space, bridging the gaps between the two most widely used atlases in histology and MRI. In addition, to address the spatial misalignment issues in MRI reconstruction and super-resolution, Wei et al. proposed a MisAlignment-Resistant Deep Unfolding Network (MAR-DUN)~\cite{wei2024misalignment} embedded in the tailored gradient descent module (GDM) and proximal mapping module (PMM) for multimodal MRI SR and reconstruction. By adaptively learning the spatial variations between the reference modality and the undersampled modality for alignment, their proposed method achieved the best reconstruction performance at the time, even in the presence of misalignment.

\begin{figure}[t]  
    \centering
    \includegraphics[width=0.7\textwidth]{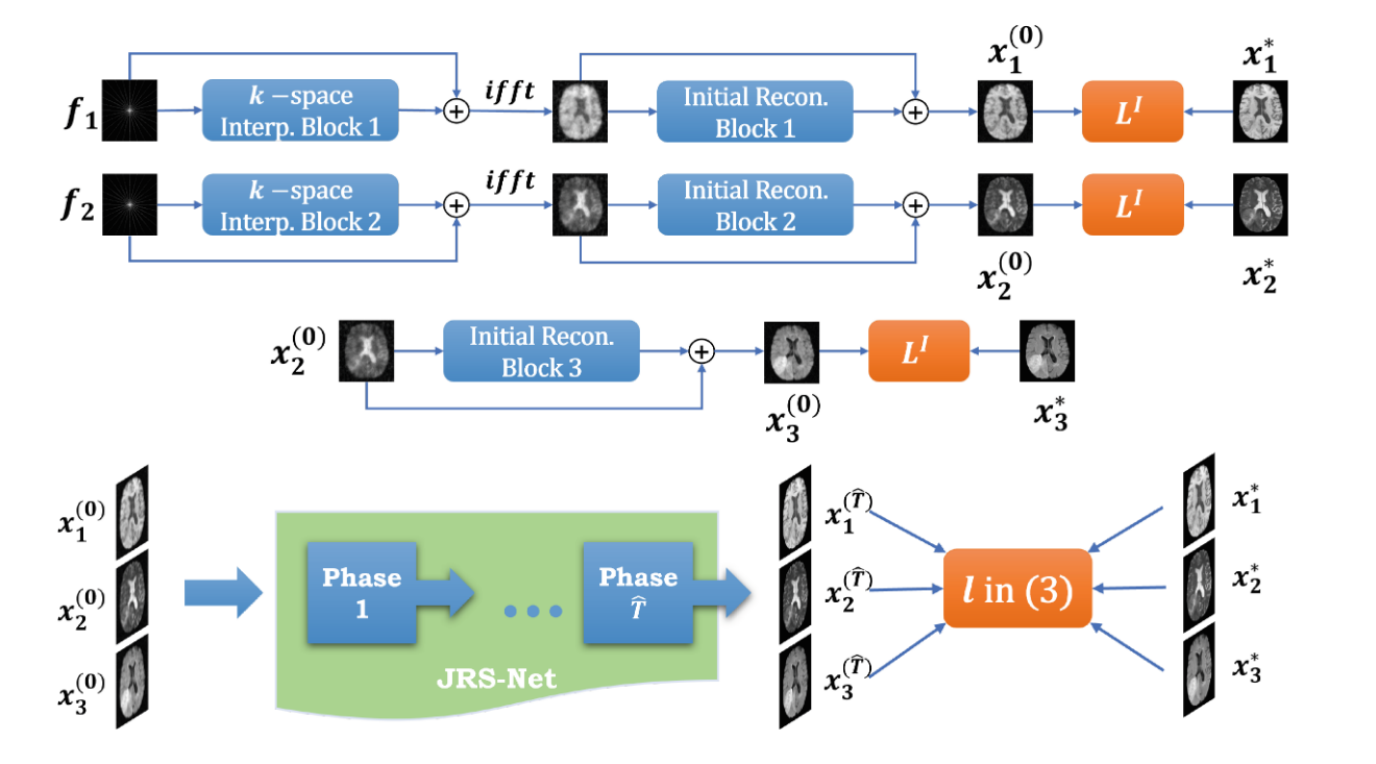}  
    \caption{The overall architecture of the proposed network of for joint multimodal MRI reconstruction and synthesis: INIT-Nets (up and middle), JRS-Net (bottom). From~\cite{bian2022learnable}.}  
    \label{fig:mutimodal}  
\end{figure}

Sometimes researchers integrate MRI reconstruction and MRI synthesis into a single process to perform both tasks simultaneously~\cite{bian2022learnable,wang2024spatial,wei2022undersampled}. For example, Due to the long acquisition time of multi-modal MRI, Bian et al. proposed a network architecture(Figure \ref{fig:mutimodal}) that combines both MRI reconstruction and MRI synthesis~\cite{bian2022learnable}. Using their proposed method, high-quality images of the source modality can be reconstructed while simultaneously synthesizing high-quality images of the target modality. In practice, this approach can significantly reduce the acquisition time of multi-modal MRI. Similarly, due to the difficulty in acquiring T2-weighted images and the susceptibility to motion artifacts, Wang et al. proposed a framework that utilizes T1-weighted images to assist in accelerating the acquisition of T2-weighted images~\cite{wang2024spatial}. This method not only improved the reconstruction results but also enabled the synthesis of T2-weighted images that are closer to real images.

The common issue of data scarcity in the medical imaging field, coupled with the increased data requirements for multi-modal reconstruction, makes federated learning a promising solution to this challenge. To address the limitation that federated learning typically requires the use of the same modality data across different hospitals, Yan et al. proposed a novel framework called Federated Consistent Regularization constrained Feature Disentanglement (Fed-CRFD)~\cite{yan2024cross}. This framework allows for the full utilization of multi-modal data from different institutions while also mitigating the domain shift problem.

\subsection{Model Training Strategies}
\subsubsection{Supervised Training Strategies}
The training strategies for deep learning-based MRI reconstruction methods can be divided into three categories: supervised, unsupervised, and semi-supervised. Among these, supervised training strategies are the most common and also the easiest to implement. In supervised training, the model is trained using paired data, meaning each undersampled input is paired with a fully sampled image as the training label. With this fully sampled label, the model can compute the difference between the reconstructed image and the label to form a loss function, which is then used as the optimization objective for training. The vast majority of MRI reconstruction methods adopt supervised training strategies.

\subsubsection{Unsupervised Training Strategies}

Due to the widespread issue of data scarcity in the field of medical imaging, researchers have been exploring ways to reduce the reliance on large amounts of annotated data. Unsupervised training strategies have garnered significant attention in recent years~\cite{li2024progressive,hu2021self,yan2023dc,zhou2022dual,zhou2023dsformer,korkmaz2022unsupervised,cole2021fast,wu2024adaptive,korkmaz2023self,cho2023improved,leynes2024scan}. 

\begin{figure}[t]  
    \centering
    \includegraphics[width=0.7\textwidth]{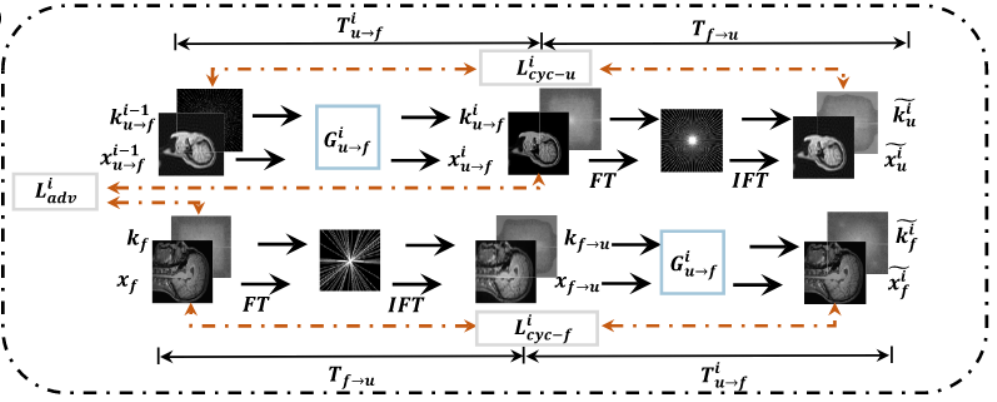}  
    \caption{In step i of PDD-GAN, unsupervised training is achieved through two cycle-consistency losses($L_{\text{cyc-f}}^i$, $L_{\text{cyc-u}}^i$), PDD-GAN~\cite{li2024progressive}.}  
    \label{fig:pdd-gan}  
\end{figure}

 Many unsupervised methods attempt to directly convert the undersampled data domain to the fully sampled data domain. However, due to the significant differences between these two domains, direct conversion often leads to a decline in reconstruction quality. To address this issue, Li et al. proposed a framework called Progressive Dual-Domain-Transfer CycleGAN (PDD-GAN)~\cite{li2024progressive}. This method breaks down the direct domain transfer problem into a multi-stage transfer process, effectively alleviating the challenges associated with domain transfer. Meanwhile, PDD-GAN introduces cycle consistency loss, eliminating the need for paired undersampled and fully sampled data, thus enabling unsupervised training and reducing reliance on paired data. It is worth noting that using cyclic consistency to achieve unsupervised training is a common idea. The implementation of PDD-GAN is shown in Figure~\ref{fig:pdd-gan}.

Unsupervised training based on K-space data partitioning is also an easy way to think of. For example, the undersampled input data is divided into subregions a and b, and then each of these subregions is reconstructed to obtain reconstruction images a and b. If the reconstruction model has been well trained, the reconstructed images a and b should theoretically be very similar. Therefore, the difference between these two images can be used as part of the loss function and as the optimization target for training. In this way, the model no longer relies on paired fully sampled images as labels, thus enabling unsupervised training. For example, Hu et al. proposed a self-supervised learning method~\cite{hu2021self}. They randomly split the undersampled k-space data of the same MRI image into two subsets, which are then input into two parallel reconstruction networks to obtain two reconstructed images. By defining the difference between these two reconstructed images as the loss, they achieved unsupervised training, thereby enhancing the model's reconstruction quality and stability. Self-supervised reconstruction only uses undersampled images for training, and because there are no fully sampled images available for reference, it often results in unnatural texture details and overly smoothed tissue regions. To address this issue, Yan et al. propose a self-supervised Deep Contrastive Siamese Network (DC-SiamNet)~\cite{yan2023dc}. Similarly, this method also employs a partitioning strategy for unsupervised training. To accelerate non-Cartesian MRI reconstruction and leverage unsupervised training strategies, Zhou et al.~\cite{zhou2022dual}proposed an MRI reconstruction method that involves partitioning both k-space and image domain data for unsupervised training. By simultaneously partitioning data in both k-space and image domains, their approach achieved higher reconstruction quality compared to traditional reconstruction methods. Many reconstruction networks use convolutional architectures, but these architectures have limited performance when capturing long-range dependencies. Additionally, to reduce the reliance on paired data for MRI reconstruction, Zhou et al. proposed a dual-domain self-supervised transformer (DSFormer)~\cite{zhou2023dsformer} to address the aforementioned issues. This method also achieves unsupervised learning through partitioning, and reconstructing each subregion individually. 

 Elizabeth K. Cole and colleagues proposed a different unsupervised training framework~\cite{cole2021fast}. This framework simulates the undersampling process of the reconstructed image generated by the generator using random undersampling masks. The resulting simulated undersampled images are then fed into the discriminator along with the real undersampled data. This encourages the generator to learn to produce reconstructed images that closely approximate the real distribution. Through this approach, the framework achieves unsupervised training and improves reconstruction quality, particularly in data-scarce situations.

Yilmaz Korkmaz et al. proposed Self-Supervised Diffusion Reconstruction (SSDiffRecon)~\cite{korkmaz2023self}, which combines diffusion models and only requires undersampled images for training, achieving an unsupervised training approach. Experiments showed that it achieved the best reconstruction quality at the time. Similarly, Cho et al. also combined diffusion models and proposed zero-MIRID~\cite{cho2023improved}(zero-shot self-supervised learning of Multi-shot Image Reconstruction for Improved Diffusion MRI). This method is trained in an unsupervised manner, and experiments demonstrated that it achieved better reconstruction quality compared to the state-of-the-art parallel imaging methods at the time.

\subsubsection{Semi-supervised Training Strategies}

\begin{figure}[t]  
    \centering
    \includegraphics[width=0.6\textwidth]{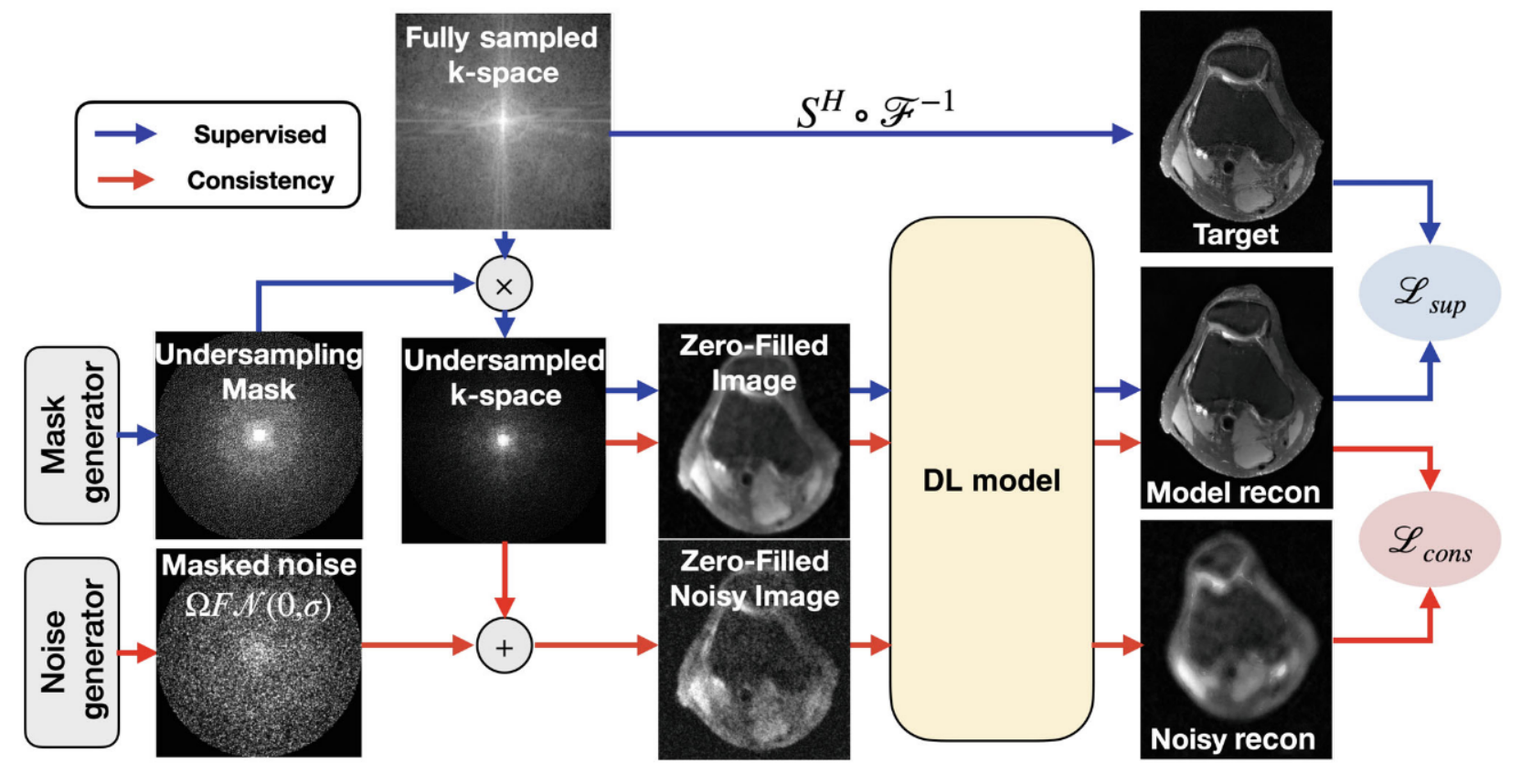}  
    \caption{The architecture of Noise2Recon~\cite{desai2023noise2recon}.}  
    \label{fig:semi}  
\end{figure}

Semi-supervised Learning is a learning method that lies between Supervised Learning and Unsupervised Learning~\cite{jiang2024towards}. In semi-supervised learning, the model is trained using a small amount of labeled data and a large amount of unlabeled data. The core idea behind this approach is that although labeled data is typically scarce and expensive to obtain, by combining unlabeled data, the model can still learn useful information, thereby improving its performance. The use of semi-supervised training strategies combines the advantages of both supervised and unsupervised learning, requiring fewer labeled data while effectively extracting valuable information from the labeled data. Arjun D. Desai et al. proposed Noise2Recon~\cite{desai2023noise2recon}, a method that can be trained in a semi-supervised manner. The architecture diagram of Noise2Recon is shown in Figure~\ref{fig:semi}. In the figure, the fully-sampled images are processed through the blue arrow stream for supervised learning, while the undersampled images are processed through the red arrow stream for unsupervised learning. This method not only performs reconstruction but also removes noise, thereby improving image quality. Even in scenarios with limited data, it is capable of producing high-quality reconstructed images.

\section{Loss Functions and Evaluation of Reconstruction Algorithms}\label{5}
\subsection{Overview of Loss Functions}

In various methods of MRI reconstruction, a range of loss functions are employed as optimization objectives during the training process. The design of an appropriate loss function plays a crucial role in improving the performance of these methods. Therefore, this section focus on introducing several commonly used loss functions in the field of MRI reconstruction and discussing their application in different reconstruction methods as well as their impact on the performance.

(1)\textbf{\(\mathcal{L}_1\) and \(\mathcal{L}_2\) losses}

In MRI reconstruction tasks, the goal is to reconstruct low-quality images into high-quality ones. During model training, a high-quality image is typically provided as a reference label. In this case, the most straightforward optimization objective is to compute the difference between the reconstructed image and the reference image, and to optimize the model by minimizing this difference. Therefore, the most common approach is to use \(\mathcal{L}_1\) or \(\mathcal{L}_2\) norms to measure the difference between the reconstructed image and the reference image, and to use them as the loss function.

\(\mathcal{L}_1\) loss calculates the total sum of the absolute differences between the pixel values of the reference image and the reconstructed image at each pixel position. The formula is given by Eq.~\ref{eq:L1}:

\begin{equation}
\mathcal{L}_1 = \sum_i \left| x_i - y_i \right|,
\label{eq:L1}
\end{equation}
where \( x_i \) and \( y_i \) represent the pixel values of the reference image and the reconstructed image at position \( i \), respectively. \(\mathcal{L}_1\) loss is more robust to noisy images because it is less sensitive to outliers.

\(\mathcal{L}_2\) loss calculates the sum of the squared differences between the pixel values of the reference image and the reconstructed image at each pixel position. The formula is given by Eq.~\ref{eq:L2}:

\begin{equation}
\mathcal{L}_2 = \sum_i \left( x_i - y_i \right)^2,
\label{eq:L2}
\end{equation}
where \( x_i \) and \( y_i \) represent the pixel values of the reference image and the reconstructed image at position \( i \), respectively. \(\mathcal{L}_2\) loss is more sensitive to larger errors because it amplifies the differences, making it suitable for improving the precision and overall quality of the image. \(\mathcal{L}_2\) loss typically encourages the model to output pixel values that are closer to those of the true image.

In the SwinGAN method, Zhao et al. proposed using both \(\mathcal{L}_1\) and \(\mathcal{L}_2\) losses to optimize MRI reconstruction tasks~\cite{swingan}. Specifically, the \(\mathcal{L}_2\) loss term helps ensure robust convergence during training, while the \(\mathcal{L}_1\) loss term prevents oversmoothing of the reconstructed image, thereby maintaining image details and edges while ensuring convergence speed.

(2) \textbf{Adversarial loss }

In recent years, Generative Adversarial Networks (GANs)~\cite{gan} have been widely applied in the field of MRI reconstruction. The core mechanism of GANs involves the introduction of a generator-discriminator adversarial structure, where the competition between the two enables iterative optimization. The key loss function used in this process is the adversarial loss. In various MRI reconstruction methods, many studies have adopted adversarial loss to enhance the quality and realism of the reconstructed images.

The adversarial loss function can be expressed as Eq.~\ref{eq:ganloss}:

\begin{equation}
L_{\text{adv}} = \mathbb{E}_{x \sim p_{\text{data}}(x)}[\log D(x)] + \mathbb{E}_{z \sim p_z(z)}[\log(1 - D(G(z)))]
\label{eq:ganloss}
\end{equation}

We can interpret the GAN as a minimax game between the generator \( G \) and the discriminator \( D \), where \( D \) aims to minimize its classification error in distinguishing fake samples from real ones, thus maximizing the loss function. On the other hand, \( G \) tries to maximize the error of the discriminator network, thereby minimizing the loss function~\cite{minaee2021image}. After training, the generator learns to produce samples that are indistinguishable from real data.

The SwinGAN method proposed by Zhao et al. combines the Swin Transformer and adversarial loss, significantly improving the quality of MRI image reconstruction~\cite{swingan}. Through adversarial training between the generator and the discriminator, this method not only enhances the reconstruction accuracy but also effectively reduces the generation of artifacts. In the DLGAN method, Noor et al. propose using a Generative Adversarial Network (GAN) to effectively handle undersampling, enhancing the quality and accuracy of MRI images while reducing the scanning time~\cite{noor2024dlgan}.

(3) \textbf{Perceptual loss}

Perceptual Loss is a deep learning-based loss function typically used to optimize image reconstruction tasks. Unlike traditional pixel-wise losses (such as \(\mathcal{L}_1\) and \(\mathcal{L}_2\) losses), perceptual loss focuses on the high-level features of the image rather than simple pixel differences. It measures the similarity between images by calculating the differences in the intermediate layers of a deep neural network, with the goal of making the reconstructed image visually closer to the real image, especially in terms of details and structure. Perceptual loss often utilizes a pre-trained network (such as the VGG network) to extract high-level features of the image, and then calculates the difference between these high-level features as the loss function.

In the DAGAN method, Yang et al. used VGG as a feature extraction network to extract high-level features~\cite{yang2017dagan}, and then computed the loss based on these features, as shown in Eq.~\ref{eq:vgg}:

\begin{equation}
 \mathcal{L}_{\text{VGG}} = \frac{1}{2} \| f_{\text{vgg}}(x_t) - f_{\text{vgg}}(\hat{x}_u) \|_2^2
\label{eq:vgg}
\end{equation}

\( f_{\text{vgg}}() \) represents the feature extraction through the VGG network, \( x_t \) denotes the reference image, and \( \hat{x}_u \) denotes the reconstructed image.

After adding perceptual loss, qualitative visualization significantly shows finer reconstruction details while avoiding unrealistic jagged artifacts, resulting in a notable improvement in visual quality~\cite{yang2017dagan,ledig2017photo}. Moreover, perceptual loss helps the model better restore image details, generating more realistic and natural images, rather than merely reducing pixel-level errors.

(4) \textbf{Identical loss}

In the PDD-GAN method, to prevent the generator from making unnecessary alterations to the input, Li et al. introduced an Identical Loss~\cite{li2024progressive} (as shown in Eq.~\ref{eq:id_loss}). By incorporating this loss, the generator is encouraged to output the input as-is when processing samples from the target domain, avoiding unnecessary transformations. Furthermore, this loss helps constrain the behavior of the generator during training, reducing the likelihood of making additional changes to data that is already close to the target domain.

\begin{equation}
\mathcal{L}_{idt}(G_{u \rightarrow f}) = \mathbb{E} \left[ \left\| G_{u \rightarrow f}(k_f, x_f) - x_f \right\|_1 \right] + \mathbb{E} \left[ \left\| G_{u \rightarrow f}(k_f, x_f) - k_f \right\|_1 \right]
\label{eq:id_loss},
\end{equation}
where $k$ denotes k-space data, $x$ represents image-domain data, $f$ indicates fully-sampled data, and $G_{u \rightarrow f}$ represents the reconstruction process. By incorporating this loss, if the input is fully-sampled target domain data, the model will not make unnecessary changes, as the input already meets the required conditions and thus does not need any further modification. In scenarios where high-quality image generation is required and unnecessary transformations of the input need to be avoided, Identical Loss is an effective loss function. Therefore, it is a commonly used loss function in the field of MRI reconstruction.

(5) \textbf{Cycle Consistency loss}

Cycle Consistency is a key concept in image-to-image translation tasks, first introduced by Zhu et al. in 2017 to address the problem of Unpaired Image-to-Image Translation~\cite{zhu2017unpaired}. It is typically applied in unsupervised learning environments. Cycle consistency requires that an input image, after passing through a generator network and undergoing two transformations (e.g., from domain A to domain B, and then from domain B to domain A), can be restored to the original input image. This consistency constraint ensures that important structural information is preserved during the transformation process, avoiding the introduction of artifacts or unnecessary changes. In MRI reconstruction, it can be viewed as an image-to-image translation task, where low-quality images are transformed into high-quality images, making it highly suitable for cycle consistency applications.

In many unsupervised MRI reconstruction methods, cycle consistency loss is widely used. By introducing this loss, the model can be effectively trained without paired samples, providing an effective solution to the data scarcity problem that is prevalent in the field of medical imaging.

Cycle Consistency loss is given by Eq.~\ref{eq:cyc_loss}

\begin{equation}
\mathcal{L}_{cyc}(G, F) = \mathbb{E}_{x \sim \mathbb{P}_{data}(x)} \left[ \left\| F(G(x)) - x \right\|_1 \right] + \mathbb{E}_{y \sim \mathbb{P}_{data}(y)} \left[ \left\| G(F(y)) - y \right\|_1 \right],
\label{eq:cyc_loss}
\end{equation}
where $x$ and $y$ represent the undersampled domain and the fully sampled domain, respectively, and $G$ and $F$ represent the reconstruction process and the undersampling process, respectively.

In the PDD-GAN method, Li et al. employed cycle consistency loss to train the MRI reconstruction model in an unsupervised manner~\cite{li2024progressive}, achieving promising results. Similarly, Li et al.~\cite{li2021high} also incorporated cycle consistency loss as part of the optimization objective to enhance the reconstruction performance.

\subsection{Performance evaluation} 
\subsubsection{Evaluation based on image related metrics}
The quantitative analysis of MRI reconstruction most frequently employs evaluation metrics such as Peak Signal-to-Noise Ratio (PSNR), Structural Similarity Index (SSIM) and Mean Squared Error(MSE)~\cite{swingan}. PSNR mainly evaluates the signal-to-noise ratio between the reconstructed image and the original image, with higher PSNR values generally indicating better image quality. SSIM assesses the quality of the reconstructed image from the perspective of structural similarity, which is more aligned with human visual perception and better reflects the image's details and structural information. MSE quantifies the reconstruction error by calculating the average pixel difference, with smaller values indicating better reconstruction performance. In addition to these metrics, Root Mean Square Error (RMSE), Normalized Root Mean Square Error (NRMSE), Normalized Mean Square Error (NMSE), Relative L2 Norm Error (RLNE), Normalized Cross Correlation (NCC), and High-Frequency Error Norm (HFEN) are extensively utilized evaluation metrics across various studies. The following provides a detailed introduction to the meanings of these metrics.

(1) \textbf{Mean Squared Error(MSE)} calculates the average of the squared differences between each pixel value of the reconstructed image ($y_i$) and the reference image ($x_i$). A smaller MSE value indicates less error between the reconstructed image and the original image, suggesting better reconstruction performance. The computation of MSE is given by Eq.~\ref{eq:mse}:

\begin{equation}
MSE = \frac{1}{N} \sum_{i=1}^{N} (x_i - y_i)^2,
\label{eq:mse}  
\end{equation}
where $x_i$ represents the $i$-th pixel of the reference image (ground truth), and $y_i$ represents the $i$-th pixel of the reconstructed image, and N is the number of all pixels in the reference image (ground truth) and the reconstructed image.

(2)\textbf{Peak Signal-to-Noise Ratio (PSNR)} is primarily used to measure the difference between the reconstructed image and the original image. It evaluates image quality by calculating the ratio between the maximum possible signal and the noise within the image. Higher PSNR values typically indicate smaller pixel-wise differences between the reconstructed image and the original image, suggesting better image quality. The computation of PSNR is given by Eq.~\ref{eq:psnr}, where $MAX_I$ represents the maximum possible pixel value of the image (e.g., 255 for an 8-bit image):

\begin{equation}
PSNR = 10 \log_{10} \left( \frac{MAX_I^2}{MSE} \right)
\label{eq:psnr}  
\end{equation}

(3) \textbf{Structural Similarity Index (SSIM)} is primarily used to measure the structural similarity between the reconstructed image and the original image. Unlike traditional pixel-level metrics such as PSNR and MSE, SSIM provides a quality measure that better aligns with human visual perception by considering the image's brightness, contrast, and structural information. The computation of SSIM is given by Eq.~\ref{eq:ssim}:

\begin{equation}
SSIM(x, y) = \frac{(2\mu_x \mu_y + C_1)(2\sigma_{xy} + C_2)}{(\mu_x^2 + \mu_y^2 + C_1)(\sigma_x^2 + \sigma_y^2 + C_2)},
\label{eq:ssim}
\end{equation}
where $x$ and $y$ represent the reference image and the reconstructed image, respectively. \( \mu \) is the mean (average) value of the image, representing the brightness of the image. \( \sigma^2 \) is the variances of the image, representing the contrast of the image. \( \sigma_{xy} \) is the covariance between the reference and reconstructed images, reflecting the structural similarity. \( C_1 \) and \( C_2 \) are small constants added to stabilize the denominator and avoid division by zero.

(4) \textbf{Normalized Mean Square Error (NMSE)}, \textbf{Root Mean Square Error (RMSE)}, and \textbf{Normalized Root Mean Square Error (NRMSE)} are variations of the MSE that provide alternative ways to evaluate the reconstruction performance. These metrics are used to measure the difference between the reference and reconstructed images, with some variations in how the error is computed or normalized.

 \textbf{NMSE} is the mean square error normalized by the sum of the squares of all pixel values of the original image.  \textbf{RMSE} is the square root of MSE. \textbf{NRMSE} is the RMSE normalized by the range of pixel values (i.e., the dynamic range). This normalization ensures that the error is independent of the image intensity and can be compared across images with different intensity ranges.

The computation of these metrics is given by the following equations:

\begin{equation}
NMSE =  \sum_{i=1}^{N} (x_i - y_i)^2 / \sum_{i=1}^{N} x_i^2
\label{eq:nmse}
\end{equation}

\begin{equation}
RMSE = \sqrt{\frac{1}{N} \sum_{i=1}^{N} (x_i - y_i)^2}
\label{eq:rmse}
\end{equation}

\begin{equation}
NRMSE = \frac{RMSE}{\max(x) - \min(x)},
\label{eq:nrmse}
\end{equation}
where $x$ represents the reference image (ground truth) and $y$ represents the reconstructed image, and N is the number of all pixels in the reference image (ground truth) and the reconstructed image.

(5) \textbf{Relative L2 Norm Error (RLNE)} is primarily used to measure the difference between the reconstructed image and the original image~\cite{ma2024mri,qu2012undersampled}. It is a normalized version of the L2 norm, which makes the error measurement independent of the image's intensity or scale. The L2 norm typically calculates the square root of the sum of squared differences between the pixel values of the original and reconstructed images. RLNE normalizes this L2 error by dividing it by the L2 norm of the original image. This normalization helps to eliminate the influence of image intensity, allowing for fairer comparisons between images of different brightness or scale. The computation of RLNE is given by Eq.~\ref{eq:rlne}:

\begin{equation}
RLNE = \frac{\| x - y \|_2}{\| x \|_2},
\label{eq:rlne}
\end{equation}
where \( x \) is the reference image, and \( y \) is the reconstructed image.

(6) \textbf{Normalized Cross Correlation (NCC)~\cite{cordero2022fetal}} is a commonly used image quality assessment metric, particularly suitable for evaluating the similarity between the reconstructed image and the original image. NCC measures the relative degree of matching between two images, with a focus on their structural information, independent of variations in image brightness or contrast. By normalizing the correlation, NCC removes the influence of image intensity, making it more sensitive to structural similarity. In the context of MRI reconstruction, NCC can be used to assess the structural similarity between the reconstructed and original images, especially when there are variations in contrast or brightness between the images.

The NCC calculation formula is given by Eq.~\ref{eq:NCC}:

\begin{equation}
NCC(x, y) = \frac{\sum_i (x_i - \bar{x})(y_i - \bar{y})}{\sqrt{\sum_i (x_i - \bar{x})^2 \sum_i (y_i - \bar{y})^2}}, 
\label{eq:NCC}  
\end{equation}
where \( x_i \) and \( y_i \) represent the pixel values at position \(i\) in the reference and reconstructed images, respectively.

(7) \textbf{High-Frequency Error Norm (HFEN)} is a specialized error metric used to evaluate image reconstruction quality, particularly in medical imaging reconstruction, such as MRI reconstruction. Unlike traditional error metrics, such as Mean Squared Error (MSE), HFEN places greater emphasis on the high-frequency components of an image, which are crucial for preserving fine details, edges, and texture information.

HFEN works by transforming the image into the frequency domain to assess the error in the high-frequency components. High-frequency information in an image typically carries detailed structural features, such as edges and textures, which are critical in the image reconstruction process. Therefore, HFEN measures the difference in high-frequency components between the original and reconstructed images, allowing for a comparison of their structural detail preservation.

The HFEN calculation formula is given by Eq.~\ref{eq:HFEN}~\cite{hfen1,hfen2,hfen3}:

\begin{equation}
HFEN = \frac{\| LoG(x_{u\to f}) - LoG(x_f) \|_2}{\| LoG(x_f) \|_2},
\label{eq:HFEN}  
\end{equation}
where \( x_{u\to f} \) and \( x_f \) represent the reconstruction image and the reference image, respectively. \( LoG \) represents the Laplacian of the Gaussian filter, which is used to capture detailed textures and high-frequency information.
 
It is worth noting that, in addition to the aforementioned evaluation metrics, visual quality is also very important, with some studies specifically designing MRI reconstruction methods to optimize visual quality~\cite{dong2022invertible}.

\subsubsection{Evaluation based on downstream tasks}

The primary goal of obtaining high-quality MRI images is to extract clinically relevant information that further supports the physician's diagnosis or other therapeutic activities. Therefore, acquiring high-quality MRI images in the shortest possible time is merely the first step in the imaging workflow~\cite{heckel2024deep}. MRI imaging can be viewed as an upstream task, with the purpose of obtaining high-quality images serving downstream tasks such as medical image segmentation~\cite{gu2022review,zhang2021deep}, lesion detection, and more. In many cases, the reconstructed images can serve as inputs for downstream tasks~\cite{sheng2024cascade}. In other words, the objective of MRI reconstruction is to improve the accuracy and other performance metrics of downstream tasks. As a result, we can evaluate the quality of reconstruction by using the reconstructed images in these downstream tasks and assessing their performance.

Specifically, MRI reconstruction can typically be combined with the following three different downstream tasks: image classification, abnormality detection, and image segmentation~\cite{heckel2024deep}. Image classification refers to identifying the presence of disease in an image, abnormality detection involves using bounding boxes to accurately locate the disease in the image, and image segmentation aims to differentiate various regions in the image (such as organs, tissues, lesion areas, etc.) from the background or other tissues, allowing for clearer observation of lesions, abnormalities, or important anatomical structures in the medical image.

Currently, there is increasing attention among researchers on the integration of MRI reconstruction with downstream tasks. Some public datasets have started to focus on meeting the goals of both MRI reconstruction and downstream task training. For example, the fastMRI dataset has been expanded into the fastMRI+ dataset, which includes classification and detection bounding box annotations for knee and brain abnormalities at the slice level~\cite{zhao2022fastmri+}. By using such datasets, not only can reconstruction tasks be performed, but it can also enhance the performance of lesion detection~\cite{zhao2021end}. For example, since deep learning-based MRI reconstruction methods often focus solely on reconstruction quality while neglecting the performance of the reconstructed images in downstream tasks. To address this issue, Jeong et al. proposed a reconstruction framework optimized for downstream tasks, and demonstrated that a single MR reconstruction network can be optimized for multiple downstream tasks by deploying continual learning (MOST)~\cite{jeong2024most}. This improvement enables the MRI reconstruction network to be applicable to multiple downstream tasks.

\section{Challenges and Future directions}\label{6}

MRI reconstruction is a key area in medical imaging, with widespread applications in medical. It plays a crucial role in assisting doctors with clinical diagnosis and provides essential data for downstream tasks, such as medical image segmentation. This is especially significant in the field of medical imaging, where data is often scarce. In recent years, the application of deep learning and multimodal technologies has led to significant advancements in MRI reconstruction. However, challenges still remain, and further research is needed to drive its continued development.

\subsection{Challenges with domain discrepancies}

In the task of MRI reconstruction, the goal is to transform under-sampled data into fully-sampled data. However, due to the significant disparity between these two domains, the reconstruction outcomes are often suboptimal. Many current methods focus on studying how to perfectly achieve this domain transfer. For instance, Li et al.~\cite{li2024progressive} attempted to decompose the direct domain transfer problem into a multi-stage domain transfer problem, hoping to improve the reconstruction quality by reducing the gap between the two domains involved in the transfer. Thus, exploring methods for transitioning between these two domains is an important research direction.

\subsection{Challenges with limited datasets}

Data scarcity is a common issue in the field of medical imaging. Unlike traditional MRI reconstruction methods, deep learning-based MRI reconstruction requires large amounts of data for training. When data is insufficient, the model is prone to overfitting, which results in a trained model that cannot meet the expected performance in clinical applications. To address the data scarcity problem in MRI reconstruction, researchers have been continuously working on finding solutions. Recently, an increasing number of publicly available medical imaging datasets have been released, alleviating the data shortage to some extent. To solve this issue, many unsupervised reconstruction methods have emerged~\cite{li2024progressive,hu2021self,yan2023dc,zhou2022dual,zhou2023dsformer,korkmaz2022unsupervised,cole2021fast,wu2024adaptive,korkmaz2023self,cho2023improved,leynes2024scan}, aiming to reduce the reliance on paired data and thus ease the challenges posed by data scarcity. Therefore, unsupervised methods will be an important direction for the development of MRI reconstruction in the future.

\subsection{Challenges with computational cost}

Deep learning-based MRI reconstruction methods typically use deep network architectures, which are very deep in terms of layers. As a result, the model needs to learn a large number of parameters, leading to high demands for computational resources~\cite{wang2024computation}. The high demand for computational resources is the reason why there are relatively few 3D reconstruction methods currently. However, 3D reconstruction can fully utilize the spatial information between adjacent slices, potentially achieving higher reconstruction performance. Therefore, 3D reconstruction is a promising direction for MRI reconstruction. To make progress in 3D reconstruction, it is essential to consider how to improve computational efficiency. Many researchers are already working on improving this efficiency~\cite{ding2022mri,yi2023frequency,feng2023learning,gutierrez2021reducing,zhang2022accelerated}. For example, Sun and colleagues have significantly enhanced computational efficiency by shifting convolution operations to the frequency domain~\cite{sun2025fourier}. Therefore, improving computational efficiency is an important direction for future research in the field of MRI reconstruction.

\subsection{Challenges with Clinical Practice}

Although some methods have been developed specifically for clinical practice (such as for cardiac imaging), most approaches are based on publicly available datasets like FastMRI. Models trained on these public datasets may not achieve satisfactory reconstruction quality when applied to clinical settings, and may not meet the needs of doctors for clinical diagnosis. Thus, future MRI reconstruction methods should not only focus on enhancing performance on public datasets but also delve into exploring the reconstruction capabilities of the proposed techniques on specific clinical data.

\subsection{Challenges with Interpretability}

Interpretability has always been an important topic in the field of deep learning. Enhancing model interpretability can help doctors better understand the model's behavior, allowing them to make more accurate decisions in clinical diagnoses. Research on uncertainty assessment has also made significant progress~\cite{zach2023stable,ding2022mri,fan2023interpretable,yang2023mgdun,edupuganti2020uncertainty}. In the future, further efforts should be made to strengthen research on interpretability, which will help promote the application of MRI reconstruction methods in clinical practice.

\subsection{Challenges with Generalization}

In MRI reconstruction, generalization capability faces numerous challenges, which are primarily due to the limitations of data diversity and task complexity. The generalization capability of a model is crucial for applying these methods to clinical practice. To enhance generalization, multi-task learning can be considered. By combining tasks such as super-resolution reconstruction and medical image synthesis, information can be shared across different tasks, thereby improving the model's generalization ability. Currently, there are many MRI reconstruction methods based on multi-task learning~\cite{beirinckx2022model,lei2023decomposition,de2023simulation,kang20243d,ji2024reconstruction,corona2021variational,wang2023mhan,sangeetha2024c2,feng2021task,yang2023mgdun,wei2024misalignment}, indicating that this is a very promising research direction.

It is worth noting that most MRI reconstruction methods use predefined undersampling masks when simulating undersampling, but some studies adopt learnable undersampling masks (for instance, an undersampling network module is used to simulate the undersampling process)~\cite{zhao2024j,hong2023dual,huang2022ista}. One explanation is that there are significant differences in the data distribution between different datasets, and fixed masks may not fully leverage these differences. In contrast, learnable undersampling masks can be optimized based on the data distribution, learning how to maximize the retention of useful information during undersampling. This means that even when dealing with images acquired using different undersampling trajectories, the model is able to better retain critical information, thereby achieving high-quality reconstruction. This characteristic significantly enhances the model's generalization capability, enabling it to adapt to diverse clinical data and sampling conditions. Therefore, in the future, it is worth further exploring the potential of learnable undersampling masks.

\section{Conclusion}\label{7}

This survey provides a comprehensive review of recent literature in the field of MRI reconstruction. We analyzed various deep learning-based MRI reconstruction methods. The paper starts by introducing MRI reconstruction and traditional reconstruction methods. The core of this survey focuses on analyzing different deep-learning approaches for MRI reconstruction based on network architectures. We discuss network models, loss functions, and evaluation methods. We also dedicate a section to multimodal MRI reconstruction methods. In addition, we examine the challenges faced in MRI reconstruction based on the literature reviewed and explore potential future research directions. Finally, we hope this review 
contribute to the development of better MRI reconstruction methods by providing a comprehensive overview of the directions suggested in the literature, and we also aim to promote its clinical application.

\noindent\textbf{CRediT authorship contribution statement}

\textbf{Xiaoyan Kui:} Writing – original draft, Supervision, Funding acquisition.
\textbf{Zijie Fan:}
Writing – original draft, Writing – review \& editing,
Investigation, Data curation.
\textbf{Zexin Ji:}
Writing – review \& editing, Methodology, Data curation. 
\textbf{Qinsong Li:}
Methodology, Supervision. 
\textbf{Chengtao Liu:}
Data curation
\textbf{Weixin Si:}
Supervision.
\textbf{Beiji Zou:}
Supervision.

\noindent\textbf{Declaration of competing interest}

The authors declare that they have no known competing financial interests or personal relationships that could have appeared to influence the work reported in this paper.

\noindent\textbf{Acknowledgements}

This work is supported by the National Natural Science Foundation of China (Nos. U22A2034, 62177047), High Caliber Foreign Experts Introduction Plan funded by MOST, Funded by Key Research and Development Programs of Department of Science and Technology of Hunan Province (No.2024JK2135) the Scientific Research Fund of Hunan Provincial Education Department (No. 24A0018), Major Program from Xiangjiang Laboratory under Grant 23XJ02005, and Central South University Research Programme of Advanced Interdisciplinary Studies (No.2023QYJC\\020).

\bibliography{mybibfile}

\begin{thebibliography}{100}
\expandafter\ifx\csname url\endcsname\relax
  \def\url#1{\texttt{#1}}\fi
\expandafter\ifx\csname urlprefix\endcsname\relax\def\urlprefix{URL }\fi
\expandafter\ifx\csname href\endcsname\relax
  \def\href#1#2{#2} \def\path#1{#1}\fi

\bibitem{zeng2021review}
G.~Zeng, Y.~Guo, J.~Zhan, Z.~Wang, Z.~Lai, X.~Du, X.~Qu, D.~Guo, A review on deep learning mri reconstruction without fully sampled k-space, BMC Medical Imaging 21~(1) (2021) 195.

\bibitem{guo2023reconformer}
P.~Guo, Y.~Mei, J.~Zhou, S.~Jiang, V.~M. Patel, Reconformer: Accelerated mri reconstruction using recurrent transformer, IEEE transactions on medical imaging.

\bibitem{li2024progressive}
B.~Li, Z.~Wang, Z.~Yang, W.~Xia, Y.~Zhang, Progressive dual-domain-transfer cyclegan for unsupervised mri reconstruction, Neurocomputing 563 (2024) 126934.

\bibitem{yi2023frequency}
Q.~Yi, F.~Fang, G.~Zhang, T.~Zeng, Frequency learning via multi-scale fourier transformer for mri reconstruction, IEEE Journal of Biomedical and Health Informatics 27~(11) (2023) 5506--5517.

\bibitem{elmas2022federated}
G.~Elmas, S.~U. Dar, Y.~Korkmaz, E.~Ceyani, B.~Susam, M.~Ozbey, S.~Avestimehr, T.~{\c{C}}ukur, Federated learning of generative image priors for mri reconstruction, IEEE Transactions on Medical Imaging 42~(7) (2022) 1996--2009.

\bibitem{sun2025fourier}
H.~Sun, Y.~Li, Z.~Li, R.~Yang, Z.~Xu, J.~Dou, H.~Qi, H.~Chen, Fourier convolution block with global receptive field for mri reconstruction, Medical Image Analysis 99 (2025) 103349.

\bibitem{xuan2022multimodal}
K.~Xuan, L.~Xiang, X.~Huang, L.~Zhang, S.~Liao, D.~Shen, Q.~Wang, Multimodal mri reconstruction assisted with spatial alignment network, IEEE Transactions on Medical Imaging 41~(9) (2022) 2499--2509.

\bibitem{arvinte2021deep}
M.~Arvinte, S.~Vishwanath, A.~H. Tewfik, J.~I. Tamir, Deep j-sense: Accelerated mri reconstruction via unrolled alternating optimization, in: International conference on medical image computing and computer-assisted intervention, Springer, 2021, pp. 350--360.

\bibitem{guo2021over}
P.~Guo, J.~M.~J. Valanarasu, P.~Wang, J.~Zhou, S.~Jiang, V.~M. Patel, Over-and-under complete convolutional rnn for mri reconstruction, in: Medical Image Computing and Computer Assisted Intervention--MICCAI 2021: 24th International Conference, Strasbourg, France, September 27--October 1, 2021, Proceedings, Part VI 24, Springer, 2021, pp. 13--23.

\bibitem{hu2021learning}
S.~Hu, N.~Pezzotti, M.~Welling, Learning to predict error for mri reconstruction, in: Medical Image Computing and Computer Assisted Intervention--MICCAI 2021: 24th International Conference, Strasbourg, France, September 27--October 1, 2021, Proceedings, Part III 24, Springer, 2021, pp. 604--613.

\bibitem{liu2021universal}
X.~Liu, J.~Wang, F.~Liu, S.~K. Zhou, Universal undersampled mri reconstruction, in: Medical Image Computing and Computer Assisted Intervention--MICCAI 2021: 24th International Conference, Strasbourg, France, September 27--October 1, 2021, Proceedings, Part VI 24, Springer, 2021, pp. 211--221.

\bibitem{murugesan2021deep}
B.~Murugesan, S.~Ramanarayanan, S.~Vijayarangan, K.~Ram, N.~R. Jagannathan, M.~Sivaprakasam, A deep cascade of ensemble of dual domain networks with gradient-based t1 assistance and perceptual refinement for fast mri reconstruction, Computerized Medical Imaging and Graphics 91 (2021) 101942.

\bibitem{narnhofer2021bayesian}
D.~Narnhofer, A.~Effland, E.~Kobler, K.~Hammernik, F.~Knoll, T.~Pock, Bayesian uncertainty estimation of learned variational mri reconstruction, IEEE transactions on medical imaging 41~(2) (2021) 279--291.

\bibitem{jun2021joint}
Y.~Jun, H.~Shin, T.~Eo, D.~Hwang, Joint deep model-based mr image and coil sensitivity reconstruction network (joint-icnet) for fast mri, in: Proceedings of the IEEE/CVF conference on computer vision and pattern recognition, 2021, pp. 5270--5279.

\bibitem{li2021multimodal}
X.-X. Li, Z.~Chen, X.-J. Lou, J.~Yang, Y.~Chen, D.~Shen, Multimodal mri acceleration via deep cascading networks with peer-layer-wise dense connections, in: Medical Image Computing and Computer Assisted Intervention--MICCAI 2021: 24th International Conference, Strasbourg, France, September 27--October 1, 2021, Proceedings, Part VI 24, Springer, 2021, pp. 329--339.

\bibitem{nitski2020cdf}
O.~Nitski, S.~Nag, C.~McIntosh, B.~Wang, Cdf-net: Cross-domain fusion network for accelerated mri reconstruction, in: International Conference on Medical Image Computing and Computer-Assisted Intervention, Springer, 2020, pp. 421--430.

\bibitem{chen2024fefa}
X.~Chen, L.~Ma, S.~Ying, D.~Shen, T.~Zeng, Fefa: Frequency enhanced multi-modal mri reconstruction with deep feature alignment, IEEE Journal of Biomedical and Health Informatics.

\bibitem{wang2022b}
G.~Wang, T.~Luo, J.-F. Nielsen, D.~C. Noll, J.~A. Fessler, B-spline parameterized joint optimization of reconstruction and k-space trajectories (bjork) for accelerated 2d mri, IEEE Transactions on Medical Imaging 41~(9) (2022) 2318--2330.

\bibitem{chen2022pyramid}
E.~Z. Chen, P.~Wang, X.~Chen, T.~Chen, S.~Sun, Pyramid convolutional rnn for mri image reconstruction, IEEE Transactions on Medical Imaging 41~(8) (2022) 2033--2047.

\bibitem{korkmaz2022unsupervised}
Y.~Korkmaz, S.~U. Dar, M.~Yurt, M.~{\"O}zbey, T.~Cukur, Unsupervised mri reconstruction via zero-shot learned adversarial transformers, IEEE Transactions on Medical Imaging 41~(7) (2022) 1747--1763.

\bibitem{wang2023dsmenet}
Y.~Wang, Y.~Pang, C.~Tong, Dsmenet: Detail and structure mutually enhancing network for under-sampled mri reconstruction, Computers in Biology and Medicine 154 (2023) 106204.

\bibitem{ramzi2022nc}
Z.~Ramzi, G.~Chaithya, J.-L. Starck, P.~Ciuciu, Nc-pdnet: A density-compensated unrolled network for 2d and 3d non-cartesian mri reconstruction, IEEE Transactions on Medical Imaging 41~(7) (2022) 1625--1638.

\bibitem{rasti2023plug}
A.~Rasti-Meymandi, A.~Ghaffari, E.~Fatemizadeh, Plug and play augmented hqs: Convergence analysis and its application in mri reconstruction, Neurocomputing 518 (2023) 1--14.

\bibitem{feng2021deep}
C.-M. Feng, H.~Fu, T.~Zhou, Y.~Xu, L.~Shao, D.~Zhang, Deep multi-modal aggregation network for mr image reconstruction with auxiliary modality, arXiv preprint arXiv:2110.08080.

\bibitem{aghabiglou2022deep}
A.~Aghabiglou, E.~M. Eksioglu, Deep unfolding architecture for mri reconstruction enhanced by adaptive noise maps, Biomedical Signal Processing and Control 78 (2022) 104016.

\bibitem{dong2022invertible}
S.~Dong, E.~Z. Chen, L.~Zhao, X.~Chen, Y.~Liu, T.~Chen, S.~Sun, Invertible sharpening network for mri reconstruction enhancement, in: International Conference on Medical Image Computing and Computer-Assisted Intervention, Springer, 2022, pp. 582--592.

\bibitem{liu2022undersampled}
X.~Liu, J.~Wang, C.~Peng, S.~S. Chandra, F.~Liu, S.~K. Zhou, Undersampled mri reconstruction with side information-guided normalisation, in: International Conference on Medical Image Computing and Computer-Assisted Intervention, Springer, 2022, pp. 323--333.

\bibitem{zhou2023rnlfnet}
L.~Zhou, M.~Zhu, D.~Xiong, L.~Ouyang, Y.~Ouyang, Z.~Chen, X.~Zhang, Rnlfnet: Residual non-local fourier network for undersampled mri reconstruction, Biomedical Signal Processing and Control 83 (2023) 104632.

\bibitem{gungor2023adaptive}
A.~G{\"u}ng{\"o}r, S.~U. Dar, {\c{S}}.~{\"O}zt{\"u}rk, Y.~Korkmaz, H.~A. Bedel, G.~Elmas, M.~Ozbey, T.~{\c{C}}ukur, Adaptive diffusion priors for accelerated mri reconstruction, Medical image analysis 88 (2023) 102872.

\bibitem{zheng2023fast}
W.-H. Zheng, X.-X. Li, H.~Hu, Q.~Zhou, Q.~Wang, Fast mri reconstruction via boosting filter diversity of deep cascading networks, in: 2023 IEEE International Conference on Bioinformatics and Biomedicine (BIBM), IEEE, 2023, pp. 2421--2426.

\bibitem{sun2023joint}
K.~Sun, Q.~Wang, D.~Shen, Joint cross-attention network with deep modality prior for fast mri reconstruction, IEEE Transactions on Medical Imaging 43~(1) (2023) 558--569.

\bibitem{ekanayake2024mcstra}
M.~Ekanayake, K.~Pawar, M.~Harandi, G.~Egan, Z.~Chen, Mcstra: A multi-branch cascaded swin transformer for point spread function-guided robust mri reconstruction, Computers in Biology and Medicine 168 (2024) 107775.

\bibitem{dar2023parallel}
S.~U.~H. Dar, {\c{S}}.~{\"O}zt{\"u}rk, M.~{\"O}zbey, K.~K. Oguz, T.~{\c{C}}ukur, Parallel-stream fusion of scan-specific and scan-general priors for learning deep mri reconstruction in low-data regimes, Computers in Biology and Medicine 167 (2023) 107610.

\bibitem{korkmaz2023self}
Y.~Korkmaz, T.~Cukur, V.~M. Patel, Self-supervised mri reconstruction with unrolled diffusion models, in: International Conference on Medical Image Computing and Computer-Assisted Intervention, Springer, 2023, pp. 491--501.

\bibitem{ozturkler2023smrd}
B.~Ozturkler, C.~Liu, B.~Eckart, M.~Mardani, J.~Song, J.~Kautz, Smrd: Sure-based robust mri reconstruction with diffusion models, in: International Conference on Medical Image Computing and Computer-Assisted Intervention, Springer, 2023, pp. 199--209.

\bibitem{zach2023stable}
M.~Zach, F.~Knoll, T.~Pock, Stable deep mri reconstruction using generative priors, IEEE Transactions on Medical Imaging 42~(12) (2023) 3817--3832.

\bibitem{chen2024accelerated}
Q.~Chen, X.~Xing, Z.~Chen, Z.~Xiong, Accelerated multi-contrast mri reconstruction via frequency and spatial mutual learning, in: International Conference on Medical Image Computing and Computer-Assisted Intervention, Springer, 2024, pp. 56--66.

\bibitem{wu2024adaptive}
Z.~Wu, X.~Li, Adaptive knowledge distillation for high-quality unsupervised mri reconstruction with model-driven priors, IEEE Journal of Biomedical and Health Informatics.

\bibitem{zhao2024center}
J.~Zhao, S.~Li, Center-to-edge denoising diffusion probabilistic models with cross-domain attention for undersampled mri reconstruction, in: International Conference on Medical Image Computing and Computer-Assisted Intervention, Springer, 2024, pp. 171--180.

\bibitem{huang2024noise}
S.~Huang, G.~Luo, X.~Wang, Z.~Chen, Y.~Wang, H.~Yang, P.-A. Heng, L.~Zhang, M.~Lyu, Noise level adaptive diffusion model for robust reconstruction of accelerated mri, in: International Conference on Medical Image Computing and Computer-Assisted Intervention, Springer, 2024, pp. 498--508.

\bibitem{wang2024spatial}
Q.~Wang, Z.~Wen, J.~Shi, Q.~Wang, D.~Shen, S.~Ying, Spatial and modal optimal transport for fast cross-modal mri reconstruction, IEEE Transactions on Medical Imaging.

\bibitem{yan2024cross}
Y.~Yan, H.~Wang, Y.~Huang, N.~He, L.~Zhu, Y.~Xu, Y.~Li, Y.~Zheng, Cross-modal vertical federated learning for mri reconstruction, IEEE Journal of Biomedical and Health Informatics.

\bibitem{chen2024multi}
J.~Chen, F.~Wu, J.~Zheng, Multi-contrast mri reconstruction via information-growth holistic unfolding network, IEEE Transactions on Instrumentation and Measurement.

\bibitem{ekanayake2025cl}
M.~Ekanayake, Z.~Chen, M.~Harandi, G.~Egan, Z.~Chen, Cl-mri: self-supervised contrastive learning to improve the accuracy of undersampled mri reconstruction, Biomedical Signal Processing and Control 100 (2025) 107185.

\bibitem{sriram2020grappanet}
A.~Sriram, J.~Zbontar, T.~Murrell, C.~L. Zitnick, A.~Defazio, D.~K. Sodickson, Grappanet: Combining parallel imaging with deep learning for multi-coil mri reconstruction, in: Proceedings of the IEEE/CVF Conference on Computer Vision and Pattern Recognition, 2020, pp. 14315--14322.

\bibitem{jeong2024most}
H.~Jeong, S.~Y. Chun, J.~Lee, Most: Mr reconstruction optimization for multiple downstream tasks via continual learning, arXiv preprint arXiv:2409.10394.

\bibitem{swingan}
X.~Zhao, T.~Yang, B.~Li, X.~Zhang, Swingan: A dual-domain swin transformer-based generative adversarial network for mri reconstruction, Computers in Biology and Medicine 153 (2023) 106513.

\bibitem{noor2024dlgan}
R.~Noor, A.~Wahid, S.~U. Bazai, A.~Khan, M.~Fang, M.~Syam, U.~A. Bhatti, Y.~Y. Ghadi, Dlgan: Undersampled mri reconstruction using deep learning based generative adversarial network, Biomedical Signal Processing and Control 93 (2024) 106218.

\bibitem{sheng2024cascade}
J.~Sheng, X.~Yang, Q.~Zhang, P.~Huang, H.~Huang, Q.~Zhang, H.~Zhu, Cascade dual-domain swin-conv-unet for mri reconstruction, Biomedical Signal Processing and Control 96 (2024) 106623.

\bibitem{feng2021task}
C.-M. Feng, Y.~Yan, H.~Fu, L.~Chen, Y.~Xu, Task transformer network for joint mri reconstruction and super-resolution, in: Medical Image Computing and Computer Assisted Intervention--MICCAI 2021: 24th International Conference, Strasbourg, France, September 27--October 1, 2021, Proceedings, Part VI 24, Springer, 2021, pp. 307--317.

\bibitem{hu2021self}
C.~Hu, C.~Li, H.~Wang, Q.~Liu, H.~Zheng, S.~Wang, Self-supervised learning for mri reconstruction with a parallel network training framework, in: Medical Image Computing and Computer Assisted Intervention--MICCAI 2021: 24th International Conference, Strasbourg, France, September 27--October 1, 2021, Proceedings, Part VI 24, Springer, 2021, pp. 382--391.

\bibitem{zhou2023dsformer}
B.~Zhou, N.~Dey, J.~Schlemper, S.~S.~M. Salehi, C.~Liu, J.~S. Duncan, M.~Sofka, Dsformer: A dual-domain self-supervised transformer for accelerated multi-contrast mri reconstruction, in: Proceedings of the IEEE/CVF winter conference on applications of computer vision, 2023, pp. 4966--4975.

\bibitem{kang20243d}
L.~Kang, B.~Tang, J.~Huang, J.~Li, 3d-mri super-resolution reconstruction using multi-modality based on multi-resolution cnn, Computer Methods and Programs in Biomedicine 248 (2024) 108110.

\bibitem{chen2022accelerating}
Z.-J. Chen, T.~Xing, Q.~Zhou, H.~Hu, X.-X. Li, Accelerating deeply cascaded convolutional networks for mri reconstruction via pixel-unshuffle caused feature squeezing, in: 2022 IEEE International Conference on Bioinformatics and Biomedicine (BIBM), IEEE, 2022, pp. 1505--1510.

\bibitem{wei2022undersampled}
H.~Wei, Z.~Li, S.~Wang, R.~Li, Undersampled multi-contrast mri reconstruction based on double-domain generative adversarial network, IEEE Journal of Biomedical and Health Informatics 26~(9) (2022) 4371--4377.

\bibitem{yan2023dc}
Y.~Yan, T.~Yang, X.~Zhao, C.~Jiao, A.~Yang, J.~Miao, Dc-siamnet: Deep contrastive siamese network for self-supervised mri reconstruction, Computers in Biology and Medicine 167 (2023) 107619.

\bibitem{lei2023decomposition}
P.~Lei, F.~Fang, G.~Zhang, T.~Zeng, Decomposition-based variational network for multi-contrast mri super-resolution and reconstruction, in: Proceedings of the IEEE/CVF International Conference on Computer Vision, 2023, pp. 21296--21306.

\bibitem{yang2023mgdun}
G.~Yang, L.~Zhang, A.~Liu, X.~Fu, X.~Chen, R.~Wang, Mgdun: an interpretable network for multi-contrast mri image super-resolution reconstruction, Computers in Biology and Medicine 167 (2023) 107605.

\bibitem{casamitjana2022robust}
A.~Casamitjana, M.~Lorenzi, S.~Ferraris, L.~Peter, M.~Modat, A.~Stevens, B.~Fischl, T.~Vercauteren, J.~E. Iglesias, Robust joint registration of multiple stains and mri for multimodal 3d histology reconstruction: Application to the allen human brain atlas, Medical image analysis 75 (2022) 102265.

\bibitem{bian2022learnable}
W.~Bian, Q.~Zhang, X.~Ye, Y.~Chen, A learnable variational model for joint multimodal mri reconstruction and synthesis, in: International Conference on Medical Image Computing and Computer-Assisted Intervention, Springer, 2022, pp. 354--364.

\bibitem{ding2022mri}
Q.~Ding, X.~Zhang, Mri reconstruction by completing under-sampled k-space data with learnable fourier interpolation, in: International Conference on Medical Image Computing and Computer-Assisted Intervention, Springer, 2022, pp. 676--685.

\bibitem{mittal20243d}
R.~Mittal, V.~Malik, G.~Singla, A.~Kaur, M.~Singh, A.~Mittal, 3d reconstruction of brain tumors from 2d mri scans: An improved marching cube algorithm, Biomedical Signal Processing and Control 91 (2024) 105901.

\bibitem{sangeetha2024c2}
G.~Sangeetha, G.~Vadivu, C2 log-gan: Concave convex and local global attention based generative adversarial network for super resolution mri reconstruction, Biomedical Signal Processing and Control 96 (2024) 106546.

\bibitem{wei2024misalignment}
J.~Wei, G.~Yang, Z.~Wang, Y.~Liu, A.~Liu, X.~Chen, Misalignment-resistant deep unfolding network for multi-modal mri super-resolution and reconstruction, Knowledge-Based Systems 296 (2024) 111866.

\bibitem{griswold2002generalized}
M.~A. Griswold, P.~M. Jakob, R.~M. Heidemann, M.~Nittka, V.~Jellus, J.~Wang, B.~Kiefer, A.~Haase, Generalized autocalibrating partially parallel acquisitions (grappa), Magnetic Resonance in Medicine: An Official Journal of the International Society for Magnetic Resonance in Medicine 47~(6) (2002) 1202--1210.

\bibitem{ullah2018qr}
I.~Ullah, H.~Nisar, H.~Raza, M.~Qasim, O.~Inam, H.~Omer, Qr-decomposition based sense reconstruction using parallel architecture, Computers in biology and medicine 95 (2018) 1--12.

\bibitem{inam2022gpu}
O.~Inam, M.~Qureshi, Z.~Laraib, H.~Akram, H.~Omer, Gpu accelerated cartesian grappa reconstruction using cuda, Journal of Magnetic Resonance 337 (2022) 107175.

\bibitem{lang2023undersampled}
J.~Lang, C.~Zhang, D.~Zhu, Undersampled mri reconstruction based on spectral graph wavelet transform, Computers in Biology and Medicine 157 (2023) 106780.

\bibitem{lustig2008compressed}
M.~Lustig, D.~L. Donoho, J.~M. Santos, J.~M. Pauly, Compressed sensing mri, IEEE signal processing magazine 25~(2) (2008) 72--82.

\bibitem{zhang2018ista}
J.~Zhang, B.~Ghanem, Ista-net: Interpretable optimization-inspired deep network for image compressive sensing, in: Proceedings of the IEEE conference on computer vision and pattern recognition, 2018, pp. 1828--1837.

\bibitem{huang2022ista}
W.~Huang, C.~Cao, S.~Hong, X.~Gao, Ista-based adaptive sparse sampling network for compressive sensing mri reconstruction, in: 2022 IEEE International Conference on Bioinformatics and Biomedicine (BIBM), IEEE, 2022, pp. 999--1004.

\bibitem{ronneberger2015u}
O.~Ronneberger, P.~Fischer, T.~Brox, U-net: Convolutional networks for biomedical image segmentation, in: Medical image computing and computer-assisted intervention--MICCAI 2015: 18th international conference, Munich, Germany, October 5-9, 2015, proceedings, part III 18, Springer, 2015, pp. 234--241.

\bibitem{gan}
I.~Goodfellow, J.~Pouget-Abadie, M.~Mirza, B.~Xu, D.~Warde-Farley, S.~Ozair, A.~Courville, Y.~Bengio, Generative adversarial nets, Advances in neural information processing systems 27.

\bibitem{zhu2017unpaired}
J.-Y. Zhu, T.~Park, P.~Isola, A.~A. Efros, Unpaired image-to-image translation using cycle-consistent adversarial networks, in: Proceedings of the IEEE international conference on computer vision, 2017, pp. 2223--2232.

\bibitem{vaswani2017attention}
A.~Vaswani, Attention is all you need, Advances in Neural Information Processing Systems.

\bibitem{guo2022attention}
M.-H. Guo, T.-X. Xu, J.-J. Liu, Z.-N. Liu, P.-T. Jiang, T.-J. Mu, S.-H. Zhang, R.~R. Martin, M.-M. Cheng, S.-M. Hu, Attention mechanisms in computer vision: A survey, Computational visual media 8~(3) (2022) 331--368.

\bibitem{zhang2023attention}
Z.~Zhang, B.~Wang, Z.~Yu, F.~Zhao, Attention guided enhancement network for weakly supervised semantic segmentation, Chinese Journal of Electronics 32~(4) (2023) 896--907.

\bibitem{dosovitskiy2020image}
A.~Dosovitskiy, An image is worth 16x16 words: Transformers for image recognition at scale, arXiv preprint arXiv:2010.11929.

\bibitem{liu2021swin}
Z.~Liu, Y.~Lin, Y.~Cao, H.~Hu, Y.~Wei, Z.~Zhang, S.~Lin, B.~Guo, Swin transformer: Hierarchical vision transformer using shifted windows, in: Proceedings of the IEEE/CVF international conference on computer vision, 2021, pp. 10012--10022.

\bibitem{ho2020denoising}
J.~Ho, A.~Jain, P.~Abbeel, Denoising diffusion probabilistic models, Advances in neural information processing systems 33 (2020) 6840--6851.

\bibitem{dayarathna2024deep}
S.~Dayarathna, K.~T. Islam, S.~Uribe, G.~Yang, M.~Hayat, Z.~Chen, Deep learning based synthesis of mri, ct and pet: Review and analysis, Medical image analysis 92 (2024) 103046.

\bibitem{peng2022towards}
C.~Peng, P.~Guo, S.~K. Zhou, V.~M. Patel, R.~Chellappa, Towards performant and reliable undersampled mr reconstruction via diffusion model sampling, in: International Conference on Medical Image Computing and Computer-Assisted Intervention, Springer, 2022, pp. 623--633.

\bibitem{gregor2010learning}
K.~Gregor, Y.~LeCun, Learning fast approximations of sparse coding, in: Proceedings of the 27th international conference on international conference on machine learning, 2010, pp. 399--406.

\bibitem{hammernik2018learning}
K.~Hammernik, T.~Klatzer, E.~Kobler, M.~P. Recht, D.~K. Sodickson, T.~Pock, F.~Knoll, Learning a variational network for reconstruction of accelerated mri data, Magnetic resonance in medicine 79~(6) (2018) 3055--3071.

\bibitem{aggarwal2018modl}
H.~K. Aggarwal, M.~P. Mani, M.~Jacob, Modl: Model-based deep learning architecture for inverse problems, IEEE transactions on medical imaging 38~(2) (2018) 394--405.

\bibitem{sun2016deep}
J.~Sun, H.~Li, Z.~Xu, et~al., Deep admm-net for compressive sensing mri, Advances in neural information processing systems 29.

\bibitem{schlemper2017deep}
J.~Schlemper, J.~Caballero, J.~V. Hajnal, A.~N. Price, D.~Rueckert, A deep cascade of convolutional neural networks for dynamic mr image reconstruction, IEEE transactions on Medical Imaging 37~(2) (2017) 491--503.

\bibitem{gu2023mamba}
A.~Gu, T.~Dao, Mamba: Linear-time sequence modeling with selective state spaces, arXiv preprint arXiv:2312.00752.

\bibitem{gu2021combining}
A.~Gu, I.~Johnson, K.~Goel, K.~Saab, T.~Dao, A.~Rudra, C.~R{\'e}, Combining recurrent, convolutional, and continuous-time models with linear state space layers, Advances in neural information processing systems 34 (2021) 572--585.

\bibitem{al2024ocucformer}
M.~Al~Fahim, S.~Ramanarayanan, G.~Rahul, M.~N. Gayathri, A.~Sarkar, K.~Ram, M.~Sivaprakasam, Ocucformer: An over-complete under-complete transformer network for accelerated mri reconstruction, Image and Vision Computing 150 (2024) 105228.

\bibitem{pan2024unrolled}
J.~Pan, M.~Hamdi, W.~Huang, K.~Hammernik, T.~Kuestner, D.~Rueckert, Unrolled and rapid motion-compensated reconstruction for cardiac cine mri, Medical Image Analysis 91 (2024) 103017.

\bibitem{huang2024mambamir}
J.~Huang, L.~Yang, F.~Wang, Y.~Nan, A.~I. Aviles-Rivero, C.-B. Sch{\"o}nlieb, D.~Zhang, G.~Yang, Mambamir: An arbitrary-masked mamba for joint medical image reconstruction and uncertainty estimation, arXiv preprint arXiv:2402.18451.

\bibitem{xu2023nesvor}
J.~Xu, D.~Moyer, B.~Gagoski, J.~E. Iglesias, P.~E. Grant, P.~Golland, E.~Adalsteinsson, Nesvor: implicit neural representation for slice-to-volume reconstruction in mri, IEEE transactions on medical imaging 42~(6) (2023) 1707--1719.

\bibitem{zou2022joint}
J.~Zou, Y.~Cao, Joint optimization of kt sampling pattern and reconstruction of dce mri for pharmacokinetic parameter estimation, IEEE transactions on medical imaging 41~(11) (2022) 3320--3331.

\bibitem{wang2024crnn}
B.~Wang, Y.~Lian, X.~Xiong, H.~Han, Z.~Liu, Crnn-refined spatiotemporal transformer for dynamic mri reconstruction, Computers in Biology and Medicine 182 (2024) 109133.

\bibitem{yiasemis2022recurrent}
G.~Yiasemis, J.-J. Sonke, C.~S{\'a}nchez, J.~Teuwen, Recurrent variational network: a deep learning inverse problem solver applied to the task of accelerated mri reconstruction, in: Proceedings of the IEEE/CVF conference on computer vision and pattern recognition, 2022, pp. 732--741.

\bibitem{bongratz2024neural}
F.~Bongratz, A.-M. Rickmann, C.~Wachinger, Neural deformation fields for template-based reconstruction of cortical surfaces from mri, Medical Image Analysis 93 (2024) 103093.

\bibitem{ren2022complex}
Y.~Ren, W.~Jiang, Y.~Liu, A complex-valued dual-domain dilated convolution neural network for brain mri reconstruction, in: 2022 IEEE International Conference on Bioinformatics and Biomedicine (BIBM), IEEE, 2022, pp. 1144--1149.

\bibitem{corona2021variational}
V.~Corona, A.~Aviles-Rivero, N.~Debroux, C.~Le~Guyader, C.-B. Sch{\"o}nlieb, Variational multi-task mri reconstruction: Joint reconstruction, registration and super-resolution, Medical Image Analysis 68 (2021) 101941.

\bibitem{wang2023mhan}
W.~Wang, H.~Shen, J.~Chen, F.~Xing, Mhan: Multi-stage hybrid attention network for mri reconstruction and super-resolution, Computers in Biology and Medicine 163 (2023) 107181.

\bibitem{wang2024dpfnet}
Y.~Wang, B.~Luo, Y.~Zhang, Z.~Xiao, M.~Wang, Y.~Niu, A.~K. Nandi, Dpfnet: Fast reconstruction of multi-coil mri based on dual domain parallel fusion network, IEEE Journal of Biomedical and Health Informatics.

\bibitem{yang2024attention}
Z.~Yang, D.~Shen, K.~W. Chan, J.~Huang, Attention-based multioffset deep learning reconstruction of chemical exchange saturation transfer (amo-cest) mri, IEEE Journal of Biomedical and Health Informatics.

\bibitem{huang2021dynamic}
Q.~Huang, Y.~Xian, D.~Yang, H.~Qu, J.~Yi, P.~Wu, D.~N. Metaxas, Dynamic mri reconstruction with end-to-end motion-guided network, Medical Image Analysis 68 (2021) 101901.

\bibitem{chen2023surfflow}
X.~Chen, J.~Zhao, S.~Liu, S.~Ahmad, P.-T. Yap, Surfflow: A flow-based approach for rapid and accurate cortical surface reconstruction from infant brain mri, in: International Conference on Medical Image Computing and Computer-Assisted Intervention, Springer, 2023, pp. 380--388.

\bibitem{li2024radial}
Z.~Li, S.~Li, Z.~Zhang, F.~Wang, F.~Wu, S.~Gao, Radial undersampled mri reconstruction using deep learning with mutual constraints between real and imaginary components of k-space, IEEE Journal of Biomedical and Health Informatics 28~(6) (2024) 3583--3596.

\bibitem{chen2021wavelet}
Y.~Chen, D.~Firmin, G.~Yang, Wavelet improved gan for mri reconstruction, in: Medical imaging 2021: Physics of medical imaging, Vol. 11595, SPIE, 2021, pp. 285--295.

\bibitem{lv2021pic}
J.~Lv, C.~Wang, G.~Yang, Pic-gan: a parallel imaging coupled generative adversarial network for accelerated multi-channel mri reconstruction, Diagnostics 11~(1) (2021) 61.

\bibitem{yaqub2022gan}
M.~Yaqub, F.~Jinchao, S.~Ahmed, K.~Arshid, M.~A. Bilal, M.~P. Akhter, M.~S. Zia, Gan-tl: Generative adversarial networks with transfer learning for mri reconstruction, Applied Sciences 12~(17) (2022) 8841.

\bibitem{han2021madgan}
C.~Han, L.~Rundo, K.~Murao, T.~Noguchi, Y.~Shimahara, Z.~{\'A}. Milacski, S.~Koshino, E.~Sala, H.~Nakayama, S.~Satoh, Madgan: Unsupervised medical anomaly detection gan using multiple adjacent brain mri slice reconstruction, BMC bioinformatics 22 (2021) 1--20.

\bibitem{yuan2020sara}
Z.~Yuan, M.~Jiang, Y.~Wang, B.~Wei, Y.~Li, P.~Wang, W.~Menpes-Smith, Z.~Niu, G.~Yang, Sara-gan: Self-attention and relative average discriminator based generative adversarial networks for fast compressed sensing mri reconstruction, Frontiers in Neuroinformatics 14 (2020) 611666.

\bibitem{cole2020unsupervised}
E.~K. Cole, J.~M. Pauly, S.~S. Vasanawala, F.~Ong, Unsupervised mri reconstruction with generative adversarial networks, arXiv preprint arXiv:2008.13065.

\bibitem{li2023cs}
X.~Li, H.~Zhang, H.~Yang, T.-Q. Li, Cs-mri reconstruction using an improved gan with dilated residual networks and channel attention mechanism, Sensors 23~(18) (2023) 7685.

\bibitem{zhang20213d}
H.~Zhang, Y.~Shinomiya, S.~Yoshida, 3d mri reconstruction based on 2d generative adversarial network super-resolution, Sensors 21~(9) (2021) 2978.

\bibitem{murugesan2019recon}
B.~Murugesan, S.~Vijaya~Raghavan, K.~Sarveswaran, K.~Ram, M.~Sivaprakasam, Recon-glgan: a global-local context based generative adversarial network for mri reconstruction, in: Machine Learning for Medical Image Reconstruction: Second International Workshop, MLMIR 2019, Held in Conjunction with MICCAI 2019, Shenzhen, China, October 17, 2019, Proceedings 2, Springer, 2019, pp. 3--15.

\bibitem{li2021modified}
G.~Li, J.~Lv, C.~Wang, A modified generative adversarial network using spatial and channel-wise attention for cs-mri reconstruction, IEEE Access 9 (2021) 83185--83198.

\bibitem{liu2022dbgan}
X.~Liu, H.~Du, J.~Xu, B.~Qiu, Dbgan: A dual-branch generative adversarial network for undersampled mri reconstruction, Magnetic Resonance Imaging 89 (2022) 77--91.

\bibitem{yang2017dagan}
G.~Yang, S.~Yu, H.~Dong, G.~Slabaugh, P.~L. Dragotti, X.~Ye, F.~Liu, S.~Arridge, J.~Keegan, Y.~Guo, et~al., Dagan: deep de-aliasing generative adversarial networks for fast compressed sensing mri reconstruction, IEEE transactions on medical imaging 37~(6) (2017) 1310--1321.

\bibitem{deora2020structure}
P.~Deora, B.~Vasudeva, S.~Bhattacharya, P.~M. Pradhan, Structure preserving compressive sensing mri reconstruction using generative adversarial networks, in: Proceedings of the IEEE/CVF conference on computer vision and pattern recognition workshops, 2020, pp. 522--523.

\bibitem{sood20213d}
R.~R. Sood, W.~Shao, C.~Kunder, N.~C. Teslovich, J.~B. Wang, S.~J. Soerensen, N.~Madhuripan, A.~Jawahar, J.~D. Brooks, P.~Ghanouni, et~al., 3d registration of pre-surgical prostate mri and histopathology images via super-resolution volume reconstruction, Medical image analysis 69 (2021) 101957.

\bibitem{zhou2021efficient}
W.~Zhou, H.~Du, W.~Mei, L.~Fang, Efficient structurally-strengthened generative adversarial network for mri reconstruction, Neurocomputing 422 (2021) 51--61.

\bibitem{wu2024compressed}
K.~Wu, Y.~Xia, N.~Ravikumar, A.~F. Frangi, Compressed sensing using a deep adaptive perceptual generative adversarial network for mri reconstruction from undersampled k-space data, Biomedical Signal Processing and Control 96 (2024) 106560.

\bibitem{lv2021transfer}
J.~Lv, G.~Li, X.~Tong, W.~Chen, J.~Huang, C.~Wang, G.~Yang, Transfer learning enhanced generative adversarial networks for multi-channel mri reconstruction, Computers in Biology and Medicine 134 (2021) 104504.

\bibitem{huang2022swin}
J.~Huang, Y.~Fang, Y.~Wu, H.~Wu, Z.~Gao, Y.~Li, J.~Del~Ser, J.~Xia, G.~Yang, Swin transformer for fast mri, Neurocomputing 493 (2022) 281--304.

\bibitem{huang2022fast}
J.~Huang, Y.~Wu, H.~Wu, G.~Yang, Fast mri reconstruction: How powerful transformers are?, in: 2022 44th annual international conference of the IEEE engineering in medicine \& biology society (EMBC), IEEE, 2022, pp. 2066--2070.

\bibitem{korkmaz2021deep}
Y.~Korkmaz, M.~Yurt, S.~U.~H. Dar, M.~{\"O}zbey, T.~Cukur, Deep mri reconstruction with generative vision transformers, in: Machine Learning for Medical Image Reconstruction: 4th International Workshop, MLMIR 2021, Held in Conjunction with MICCAI 2021, Strasbourg, France, October 1, 2021, Proceedings 4, Springer, 2021, pp. 54--64.

\bibitem{zhao2024diffgan}
X.~Zhao, T.~Yang, B.~Li, A.~Yang, Y.~Yan, C.~Jiao, Diffgan: an adversarial diffusion model with local transformer for mri reconstruction, Magnetic Resonance Imaging 109 (2024) 108--119.

\bibitem{hu2022trans}
D.~Hu, Y.~Zhang, J.~Zhu, Q.~Liu, Y.~Chen, Trans-net: Transformer-enhanced residual-error alternative suppression network for mri reconstruction, IEEE Transactions on Instrumentation and Measurement 71 (2022) 1--13.

\bibitem{shen2024magnetic}
G.~Shen, M.~Li, S.~Anderson, C.~W. Farris, X.~Zhang, Magnetic resonance image processing transformer for general reconstruction, arXiv preprint arXiv:2405.15098.

\bibitem{xu2023learning}
D.~Xu, H.~Liu, D.~Ruan, K.~Sheng, Learning dynamic mri reconstruction with convolutional network assisted reconstruction swin transformer, in: International Conference on Medical Image Computing and Computer-Assisted Intervention, Springer, 2023, pp. 3--13.

\bibitem{du2023transformer}
W.~Du, S.~Tian, Transformer and gan-based super-resolution reconstruction network for medical images, Tsinghua Science and Technology 29~(1) (2023) 197--206.

\bibitem{alghallabi2023accelerated}
W.~Alghallabi, A.~Dudhane, W.~Zamir, S.~Khan, F.~S. Khan, Accelerated mri reconstruction via dynamic deformable alignment based transformer, in: International Workshop on Machine Learning in Medical Imaging, Springer, 2023, pp. 104--114.

\bibitem{lin2022vision}
K.~Lin, R.~Heckel, Vision transformers enable fast and robust accelerated mri, in: International Conference on Medical Imaging with Deep Learning, PMLR, 2022, pp. 774--795.

\bibitem{korkmaz2022mri}
Y.~Korkmaz, M.~{\"O}zbey, T.~Cukur, Mri reconstruction with conditional adversarial transformers, in: International Workshop on Machine Learning for Medical Image Reconstruction, Springer, 2022, pp. 62--71.

\bibitem{ekanayake2022multi}
M.~Ekanayake, K.~Pawar, M.~Harandi, G.~Egan, Z.~Chen, Multi-branch cascaded swin transformers with attention to k-space sampling pattern for accelerated mri reconstruction, arXiv preprint arXiv:2207.08412.

\bibitem{feng2022multimodal}
C.-M. Feng, Y.~Yan, G.~Chen, Y.~Xu, Y.~Hu, L.~Shao, H.~Fu, Multimodal transformer for accelerated mr imaging, IEEE Transactions on Medical Imaging 42~(10) (2022) 2804--2816.

\bibitem{zhao2022k}
Z.~Zhao, T.~Zhang, W.~Xie, Y.~Wang, Y.~Zhang, K-space transformer for undersampled mri reconstruction, arXiv preprint arXiv:2206.06947.

\bibitem{wu2023deep}
Z.~Wu, W.~Liao, C.~Yan, M.~Zhao, G.~Liu, N.~Ma, X.~Li, Deep learning based mri reconstruction with transformer, Computer Methods and Programs in Biomedicine 233 (2023) 107452.

\bibitem{gao2022projection}
C.~Gao, S.-F. Shih, J.~P. Finn, X.~Zhong, A projection-based k-space transformer network for undersampled radial mri reconstruction with limited training subjects, in: International Conference on Medical Image Computing and Computer-Assisted Intervention, Springer, 2022, pp. 726--736.

\bibitem{pan2023global}
J.~Pan, S.~Shit, {\"O}.~Turgut, W.~Huang, H.~B. Li, N.~Stolt-Ans{\'o}, T.~K{\"u}stner, K.~Hammernik, D.~Rueckert, Global k-space interpolation for dynamic mri reconstruction using masked image modeling, in: International Conference on Medical Image Computing and Computer-Assisted Intervention, Springer, 2023, pp. 228--238.

\bibitem{feng2023learning}
C.-M. Feng, B.~Li, X.~Xu, Y.~Liu, H.~Fu, W.~Zuo, Learning federated visual prompt in null space for mri reconstruction, in: Proceedings of the IEEE/CVF Conference on Computer Vision and Pattern Recognition, 2023, pp. 8064--8073.

\bibitem{lyu2023region}
J.~Lyu, G.~Li, C.~Wang, C.~Qin, S.~Wang, Q.~Dou, J.~Qin, Region-focused multi-view transformer-based generative adversarial network for cardiac cine mri reconstruction, Medical Image Analysis 85 (2023) 102760.

\bibitem{webber2024diffusion}
G.~Webber, A.~J. Reader, Diffusion models for medical image reconstruction, BJR| Artificial Intelligence 1~(1) (2024) ubae013.

\bibitem{luo2023bayesian}
G.~Luo, M.~Blumenthal, M.~Heide, M.~Uecker, Bayesian mri reconstruction with joint uncertainty estimation using diffusion models, Magnetic Resonance in Medicine 90~(1) (2023) 295--311.

\bibitem{bian2024diffusion}
W.~Bian, A.~Jang, L.~Zhang, X.~Yang, Z.~Stewart, F.~Liu, Diffusion modeling with domain-conditioned prior guidance for accelerated mri and qmri reconstruction, IEEE Transactions on Medical Imaging.

\bibitem{chung2022score}
H.~Chung, J.~C. Ye, Score-based diffusion models for accelerated mri, Medical image analysis 80 (2022) 102479.

\bibitem{levac2023mri}
B.~Levac, A.~Jalal, K.~Ramchandran, J.~I. Tamir, Mri reconstruction with side information using diffusion models, in: 2023 57th Asilomar Conference on Signals, Systems, and Computers, IEEE, 2023, pp. 1436--1442.

\bibitem{kazerouni2023diffusion}
A.~Kazerouni, E.~K. Aghdam, M.~Heidari, R.~Azad, M.~Fayyaz, I.~Hacihaliloglu, D.~Merhof, Diffusion models in medical imaging: A comprehensive survey, Medical image analysis 88 (2023) 102846.

\bibitem{safari2024adaptive}
M.~Safari, Z.~Eidex, S.~Pan, R.~L. Qiu, X.~Yang, Adaptive self-supervised consistency-guided diffusion model for accelerated mri reconstruction, arXiv preprint arXiv:2406.15656.

\bibitem{jiang2024fast}
W.~Jiang, Z.~Xiong, F.~Liu, N.~Ye, H.~Sun, Fast controllable diffusion models for undersampled mri reconstruction, in: 2024 IEEE International Symposium on Biomedical Imaging (ISBI), IEEE, 2024, pp. 1--5.

\bibitem{geng2024dp}
M.~Geng, J.~Zhu, X.~Zhu, Q.~Liu, D.~Liang, Q.~Liu, Dp-mdm: Detail-preserving mr reconstruction via multiple diffusion models, arXiv preprint arXiv:2405.05763.

\bibitem{tan2024fetal}
J.~Tan, X.~Zhang, C.~Qing, C.~Yang, H.~Zhang, G.~Li, X.~Xu, Fetal mri reconstruction by global diffusion and consistent implicit representation, in: International Conference on Medical Image Computing and Computer-Assisted Intervention, Springer, 2024, pp. 329--339.

\bibitem{guan2024correlated}
Y.~Guan, C.~Yu, Z.~Cui, H.~Zhou, Q.~Liu, Correlated and multi-frequency diffusion modeling for highly under-sampled mri reconstruction, IEEE Transactions on Medical Imaging.

\bibitem{luo2025deep}
Y.~Luo, S.~Liu, J.~Ling, T.~Zhou, Y.~Ji, S.~Yao, G.~Yue, Deep unfolding network with adaptive sequential recovery for mri reconstruction, Biomedical Signal Processing and Control 103 (2025) 107364.

\bibitem{luo2024joint}
Y.~Luo, Y.~Cai, J.~Ling, Y.~Ji, Y.~Tie, S.~Yao, Joint edge optimization deep unfolding network for accelerated mri reconstruction, IEEE Transactions on Computational Imaging.

\bibitem{lei2023deep}
P.~Lei, F.~Fang, G.~Zhang, M.~Xu, Deep unfolding convolutional dictionary model for multi-contrast mri super-resolution and reconstruction, arXiv preprint arXiv:2309.01171.

\bibitem{zhang2025re}
H.~Zhang, Q.~Wang, J.~Sun, Z.~Wen, J.~Shi, S.~Ying, Re-visible dual-domain self-supervised deep unfolding network for mri reconstruction, arXiv preprint arXiv:2501.03737.

\bibitem{ma2024attention}
M.~Ma, H.~Gan, Attention-enhanced convolutional iterative unrolling network for under-sampled mri reconstruction, IEEE Transactions on Emerging Topics in Computational Intelligence.

\bibitem{wang2023deep}
C.~Wang, R.~Zhang, G.~Maliakal, S.~Ravishankar, B.~Wen, Deep reinforcement learning based unrolling network for mri reconstruction, in: 2023 IEEE 20th International Symposium on Biomedical Imaging (ISBI), IEEE, 2023, pp. 1--5.

\bibitem{zhang2024t2lr}
Y.~Zhang, P.~Li, Y.~Hu, T2lr-net: An unrolling network learning transformed tensor low-rank prior for dynamic mr image reconstruction, Computers in Biology and Medicine 170 (2024) 108034.

\bibitem{gunel2022scale}
B.~Gunel, A.~Sahiner, A.~D. Desai, A.~S. Chaudhari, S.~Vasanawala, M.~Pilanci, J.~Pauly, Scale-equivariant unrolled neural networks for data-efficient accelerated mri reconstruction, in: International Conference on Medical Image Computing and Computer-Assisted Intervention, Springer, 2022, pp. 737--747.

\bibitem{wang2021memory}
K.~Wang, M.~Kellman, C.~M. Sandino, K.~Zhang, S.~S. Vasanawala, J.~I. Tamir, S.~X. Yu, M.~Lustig, Memory-efficient learning for high-dimensional mri reconstruction, in: Medical Image Computing and Computer Assisted Intervention--MICCAI 2021: 24th International Conference, Strasbourg, France, September 27--October 1, 2021, Proceedings, Part VI 24, Springer, 2021, pp. 461--470.

\bibitem{fan2023interpretable}
X.~Fan, Y.~Yang, K.~Chen, J.~Zhang, K.~Dong, An interpretable mri reconstruction network with two-grid-cycle correction and geometric prior distillation, Biomedical Signal Processing and Control 84 (2023) 104821.

\bibitem{korkmaz2024mambarecon}
Y.~Korkmaz, V.~M. Patel, Mambarecon: Mri reconstruction with structured state space models, arXiv preprint arXiv:2409.12401.

\bibitem{huang2025enhancing}
J.~Huang, L.~Yang, F.~Wang, Y.~Wu, Y.~Nan, W.~Wu, C.~Wang, K.~Shi, A.~I. Aviles-Rivero, C.-B. Sch{\"o}nlieb, et~al., Enhancing global sensitivity and uncertainty quantification in medical image reconstruction with monte carlo arbitrary-masked mamba, Medical Image Analysis 99 (2025) 103334.

\bibitem{zhou2020dudornet}
B.~Zhou, S.~K. Zhou, Dudornet: learning a dual-domain recurrent network for fast mri reconstruction with deep t1 prior, in: Proceedings of the IEEE/CVF conference on computer vision and pattern recognition, 2020, pp. 4273--4282.

\bibitem{eo2018kiki}
T.~Eo, Y.~Jun, T.~Kim, J.~Jang, H.-J. Lee, D.~Hwang, Kiki-net: cross-domain convolutional neural networks for reconstructing undersampled magnetic resonance images, Magnetic resonance in medicine 80~(5) (2018) 2188--2201.

\bibitem{ran2020md}
M.~Ran, W.~Xia, Y.~Huang, Z.~Lu, P.~Bao, Y.~Liu, H.~Sun, J.~Zhou, Y.~Zhang, Md-recon-net: a parallel dual-domain convolutional neural network for compressed sensing mri, IEEE Transactions on Radiation and Plasma Medical Sciences 5~(1) (2020) 120--135.

\bibitem{liu2023diik}
Y.~Liu, Y.~Pang, X.~Liu, Y.~Liu, J.~Nie, Diik-net: A full-resolution cross-domain deep interaction convolutional neural network for mr image reconstruction, Neurocomputing 517 (2023) 213--222.

\bibitem{sun2020dual}
L.~Sun, Y.~Wu, B.~Shu, X.~Ding, C.~Cai, Y.~Huang, J.~Paisley, A dual-domain deep lattice network for rapid mri reconstruction, Neurocomputing 397 (2020) 94--107.

\bibitem{liu2022dual}
X.~Liu, Y.~Pang, R.~Jin, Y.~Liu, Z.~Wang, Dual-domain reconstruction network with v-net and k-net for fast mri, Magnetic Resonance in Medicine 88~(6) (2022) 2694--2708.

\bibitem{wang2024dct}
B.~Wang, Y.~Lian, X.~Xiong, H.~Zhou, Z.~Liu, X.~Zhou, Dct-net: Dual-domain cross-fusion transformer network for mri reconstruction, Magnetic Resonance Imaging 107 (2024) 69--79.

\bibitem{wang2019accelerated}
G.~Wang, E.~Gong, S.~Banerjee, J.~Pauly, G.~Zaharchuk, Accelerated mri reconstruction with dual-domain generative adversarial network, in: Machine Learning for Medical Image Reconstruction: Second International Workshop, MLMIR 2019, Held in Conjunction with MICCAI 2019, Shenzhen, China, October 17, 2019, Proceedings 2, Springer, 2019, pp. 47--57.

\bibitem{zhang2021dual}
Y.~Zhang, J.~Lyu, X.~Bi, A dual-task dual-domain model for blind mri reconstruction, Computerized Medical Imaging and Graphics 89 (2021) 101862.

\bibitem{zhang2025fdudoclnet}
H.~Zhang, T.~Yang, H.~Wang, J.~Fan, W.~Zhang, M.~Ji, Fdudoclnet: Fully dual-domain contrastive learning network for parallel mri reconstruction, Magnetic Resonance Imaging (2025) 110336.

\bibitem{seo2021dual}
H.~Seo, K.~M. Shin, Y.~Kyung, A dual domain network for mri reconstruction using gabor loss, in: 2021 IEEE International Conference on Image Processing (ICIP), IEEE, 2021, pp. 146--149.

\bibitem{lu2025dffki}
F.~Lu, X.~Xiao, Z.~Wang, Y.~Liu, J.~Zhou, Dffki-net: A dual-domain feature fusion deep convolutional neural network for under-sampled mr image reconstruction, Biomedical Signal Processing and Control 106 (2025) 107732.

\bibitem{groun2022novel}
N.~Groun, M.~Villalba-Orero, E.~Lara-Pezzi, E.~Valero, J.~Garicano-Mena, S.~Le~Clainche, A novel data-driven method for the analysis and reconstruction of cardiac cine mri, Computers in biology and medicine 151 (2022) 106317.

\bibitem{basit2023accelerating}
A.~Basit, O.~Inam, H.~Omer, Accelerating grappa reconstruction using soc design for real-time cardiac mri, Computers in Biology and Medicine 160 (2023) 107008.

\bibitem{gan2021ss}
W.~Gan, Y.~Hu, C.~Eldeniz, J.~Liu, Y.~Chen, H.~An, U.~S. Kamilov, Ss-jircs: Self-supervised joint image reconstruction and coil sensitivity calibration in parallel mri without ground truth, in: Proceedings of the IEEE/CVF International Conference on Computer Vision, 2021, pp. 4048--4056.

\bibitem{hashemizadehkolowri2021simultaneous}
S.~HashemizadehKolowri, R.-R. Chen, G.~Adluru, D.~C. Dean, E.~A. Wilde, A.~L. Alexander, E.~V. DiBella, Simultaneous multi-slice image reconstruction using regularized image domain split slice-grappa for diffusion mri, Medical Image Analysis 70 (2021) 102000.

\bibitem{liu2023high}
J.~Liu, C.~Qin, M.~Yaghoobi, High-fidelity mri reconstruction using adaptive spatial attention selection and deep data consistency prior, IEEE Transactions on Computational Imaging 9 (2023) 298--313.

\bibitem{cordero2022fetal}
L.~Cordero-Grande, J.~E. Ortu{\~n}o-Fisac, A.~A. Del~Hoyo, A.~Uus, M.~Deprez, A.~Santos, J.~V. Hajnal, M.~J. Ledesma-Carbayo, Fetal mri by robust deep generative prior reconstruction and diffeomorphic registration, IEEE Transactions on Medical Imaging 42~(3) (2022) 810--822.

\bibitem{he2024igroupss}
Y.~He, B.~Tu, P.~Jiang, B.~Liu, J.~Li, A.~Plaza, Igroupss-mamba: Interval group spatial-spectral mamba for hyperspectral image classification, IEEE Transactions on Geoscience and Remote Sensing.

\bibitem{li2024mambahsi}
Y.~Li, Y.~Luo, L.~Zhang, Z.~Wang, B.~Du, Mambahsi: Spatial-spectral mamba for hyperspectral image classification, IEEE Transactions on Geoscience and Remote Sensing.

\bibitem{yang2024remamber}
Y.~Yang, C.~Ma, J.~Yao, Z.~Zhong, Y.~Zhang, Y.~Wang, Remamber: Referring image segmentation with mamba twister, in: European Conference on Computer Vision, Springer, 2024, pp. 108--126.

\bibitem{xing2024segmamba}
Z.~Xing, T.~Ye, Y.~Yang, G.~Liu, L.~Zhu, Segmamba: Long-range sequential modeling mamba for 3d medical image segmentation, in: International Conference on Medical Image Computing and Computer-Assisted Intervention, Springer, 2024, pp. 578--588.

\bibitem{ji2024deform}
Z.~Ji, B.~Zou, X.~Kui, P.~Vera, S.~Ruan, Deform-mamba network for mri super-resolution, in: International Conference on Medical Image Computing and Computer-Assisted Intervention, Springer, 2024, pp. 242--252.

\bibitem{ji2025generation}
Z.~Ji, B.~Zou, X.~Kui, H.~Li, P.~Vera, S.~Ruan, Generation of super-resolution for medical image via a self-prior guided mamba network with edge-aware constraint, Pattern Recognition Letters 187 (2025) 93--99.

\bibitem{ji2025self}
Z.~Ji, B.~Zou, X.~Kui, P.~Vera, S.~Ruan, Self-prior guided mamba-unet networks for medical image super-resolution, in: International Conference on Pattern Recognition, Springer, 2025, pp. 160--174.

\bibitem{zou2024mmr}
J.~Zou, L.~Liu, Q.~Chen, S.~Wang, X.~Xing, J.~Qin, Mmr-mamba: Multi-contrast mri reconstruction with mamba and spatial-frequency information fusion, arXiv e-prints (2024) arXiv--2406.

\bibitem{li2023deep}
X.-X. Li, W.-H. Zheng, Q.~Zhou, H.~Hu, L.~Chen, Deep k-space partition-based convolutional networks for fast multimodal mri reconstruction, in: 2023 IEEE International Conference on Bioinformatics and Biomedicine (BIBM), IEEE, 2023, pp. 1280--1285.

\bibitem{luo2023effective}
Y.~Luo, M.~Wei, S.~Li, J.~Ling, G.~Xie, S.~Yao, An effective co-support guided analysis model for multi-contrast mri reconstruction, IEEE Journal of Biomedical and Health Informatics 27~(5) (2023) 2477--2488.

\bibitem{zhou2022dual}
B.~Zhou, J.~Schlemper, N.~Dey, S.~S.~M. Salehi, K.~Sheth, C.~Liu, J.~S. Duncan, M.~Sofka, Dual-domain self-supervised learning for accelerated non-cartesian mri reconstruction, Medical Image Analysis 81 (2022) 102538.

\bibitem{cole2021fast}
E.~K. Cole, F.~Ong, S.~S. Vasanawala, J.~M. Pauly, Fast unsupervised mri reconstruction without fully-sampled ground truth data using generative adversarial networks, in: Proceedings of the IEEE/CVF International Conference on Computer Vision, 2021, pp. 3988--3997.

\bibitem{cho2023improved}
J.~Cho, Y.~Jun, X.~Wang, C.~Kobayashi, B.~Bilgic, Improved multi-shot diffusion-weighted mri with zero-shot self-supervised learning reconstruction, in: International Conference on Medical Image Computing and Computer-Assisted Intervention, Springer, 2023, pp. 457--466.

\bibitem{leynes2024scan}
A.~P. Leynes, N.~Deveshwar, S.~S. Nagarajan, P.~E. Larson, Scan-specific self-supervised bayesian deep non-linear inversion for undersampled mri reconstruction, IEEE Transactions on Medical Imaging 43~(6) (2024) 2358--2369.

\bibitem{jiang2024towards}
T.~Jiang, W.~Chen, H.~Zhou, J.~He, P.~Qi, Towards semi-supervised classification of abnormal spectrum signals based on deep learning, Chinese Journal of Electronics 33~(3) (2024) 721--731.

\bibitem{desai2023noise2recon}
A.~D. Desai, B.~M. Ozturkler, C.~M. Sandino, R.~Boutin, M.~Willis, S.~Vasanawala, B.~A. Hargreaves, C.~R{\'e}, J.~M. Pauly, A.~S. Chaudhari, Noise2recon: Enabling snr-robust mri reconstruction with semi-supervised and self-supervised learning, Magnetic Resonance in Medicine 90~(5) (2023) 2052--2070.

\bibitem{minaee2021image}
S.~Minaee, Y.~Boykov, F.~Porikli, A.~Plaza, N.~Kehtarnavaz, D.~Terzopoulos, Image segmentation using deep learning: A survey, IEEE transactions on pattern analysis and machine intelligence 44~(7) (2021) 3523--3542.

\bibitem{ledig2017photo}
C.~Ledig, L.~Theis, F.~Husz{\'a}r, J.~Caballero, A.~Cunningham, A.~Acosta, A.~Aitken, A.~Tejani, J.~Totz, Z.~Wang, et~al., Photo-realistic single image super-resolution using a generative adversarial network, in: Proceedings of the IEEE conference on computer vision and pattern recognition, 2017, pp. 4681--4690.

\bibitem{li2021high}
G.~Li, J.~Lv, X.~Tong, C.~Wang, G.~Yang, High-resolution pelvic mri reconstruction using a generative adversarial network with attention and cyclic loss, IEEE Access 9 (2021) 105951--105964.

\bibitem{ma2024mri}
Q.~Ma, Z.~Lai, Z.~Wang, Y.~Qiu, H.~Zhang, X.~Qu, Mri reconstruction with enhanced self-similarity using graph convolutional network, BMC Medical Imaging 24~(1) (2024) 113.

\bibitem{qu2012undersampled}
X.~Qu, D.~Guo, B.~Ning, Y.~Hou, Y.~Lin, S.~Cai, Z.~Chen, Undersampled mri reconstruction with patch-based directional wavelets, Magnetic resonance imaging 30~(7) (2012) 964--977.

\bibitem{hfen1}
B.~Li, Z.~Wang, Z.~Yang, W.~Xia, Y.~Zhang, Progressive dual-domain-transfer cyclegan for unsupervised mri reconstruction, Neurocomputing 563 (2024) 126934.

\bibitem{hfen2}
Y.~Han, H.~Du, F.~Lam, W.~Mei, L.~Fang, Image reconstruction using analysis model prior, Computational and mathematical methods in medicine 2016~(1) (2016) 7571934.

\bibitem{hfen3}
S.~Ravishankar, Y.~Bresler, Mr image reconstruction from highly undersampled k-space data by dictionary learning, IEEE transactions on medical imaging 30~(5) (2010) 1028--1041.

\bibitem{heckel2024deep}
R.~Heckel, M.~Jacob, A.~Chaudhari, O.~Perlman, E.~Shimron, Deep learning for accelerated and robust mri reconstruction: a review, arXiv preprint arXiv:2404.15692.

\bibitem{gu2022review}
W.~Gu, S.~Bai, L.~Kong, A review on 2d instance segmentation based on deep neural networks, Image and Vision Computing 120 (2022) 104401.

\bibitem{zhang2021deep}
Y.~Zhang, D.~Sidib{\'e}, O.~Morel, F.~M{\'e}riaudeau, Deep multimodal fusion for semantic image segmentation: A survey, Image and Vision Computing 105 (2021) 104042.

\bibitem{zhao2022fastmri+}
R.~Zhao, B.~Yaman, Y.~Zhang, R.~Stewart, A.~Dixon, F.~Knoll, Z.~Huang, Y.~W. Lui, M.~S. Hansen, M.~P. Lungren, fastmri+, clinical pathology annotations for knee and brain fully sampled magnetic resonance imaging data, Scientific Data 9~(1) (2022) 152.

\bibitem{zhao2021end}
R.~Zhao, Y.~Zhang, B.~Yaman, M.~P. Lungren, M.~S. Hansen, End-to-end ai-based mri reconstruction and lesion detection pipeline for evaluation of deep learning image reconstruction, arXiv preprint arXiv:2109.11524.

\bibitem{wang2024computation}
Y.~Wang, Y.~Han, C.~Wang, S.~Song, Q.~Tian, G.~Huang, Computation-efficient deep learning for computer vision: A survey, Cybernetics and intelligence.

\bibitem{gutierrez2021reducing}
A.~Gutierrez, M.~Mullen, D.~Xiao, A.~Jang, T.~Froelich, M.~Garwood, J.~Haupt, Reducing the complexity of model-based mri reconstructions via sparsification, IEEE transactions on medical imaging 40~(9) (2021) 2477--2486.

\bibitem{zhang2022accelerated}
X.~Zhang, H.~Lu, D.~Guo, Z.~Lai, H.~Ye, X.~Peng, B.~Zhao, X.~Qu, Accelerated mri reconstruction with separable and enhanced low-rank hankel regularization, IEEE transactions on medical imaging 41~(9) (2022) 2486--2498.

\bibitem{edupuganti2020uncertainty}
V.~Edupuganti, M.~Mardani, S.~Vasanawala, J.~Pauly, Uncertainty quantification in deep mri reconstruction, IEEE Transactions on Medical Imaging 40~(1) (2020) 239--250.

\bibitem{beirinckx2022model}
Q.~Beirinckx, B.~Jeurissen, M.~Nicastro, D.~H. Poot, M.~Verhoye, A.~J. den Dekker, J.~Sijbers, Model-based super-resolution reconstruction with joint motion estimation for improved quantitative mri parameter mapping, Computerized Medical Imaging and Graphics 100 (2022) 102071.

\bibitem{de2023simulation}
P.~de~Dumast, T.~Sanchez, H.~Lajous, M.~Bach~Cuadra, Simulation-based parameter optimization for fetal brain mri super-resolution reconstruction, in: International Conference on Medical Image Computing and Computer-Assisted Intervention, Springer, 2023, pp. 336--346.

\bibitem{ji2024reconstruction}
Z.~Ji, B.~Zou, X.~Kui, Y.~Li, J.~Liu, W.~Zhao, C.~Zhu, Y.~Dai, Reconstruction-guided multi-stage network for mri super-resolution, in: Proceedings of the International Conference on Computer Vision and Deep Learning, 2024, pp. 1--5.

\bibitem{zhao2024j}
D.~Zhao, Y.~Huang, Y.~Gan, J.~Zheng, J-loucr: Joint learned optimized undersampling and constrained reconstruction for accelerated mri by reference-driven deep image prior, Biomedical Signal Processing and Control 87 (2024) 105513.

\bibitem{hong2023dual}
G.~Q. Hong, Y.~T. Wei, W.~A. Morley, M.~Wan, A.~J. Mertens, Y.~Su, H.-L.~M. Cheng, Dual-domain accelerated mri reconstruction using transformers with learning-based undersampling, Computerized Medical Imaging and Graphics 106 (2023) 102206.

\end{thebibliography}

\end{document}